\documentclass{aa}  

\usepackage{graphicx}
\usepackage{txfonts, natbib}
\usepackage{ulem}

\usepackage{tablefootnote}
\usepackage{longtable}            
\usepackage{rotating}
\usepackage{pdflscape}
\usepackage{multicol, subfigure}
\usepackage[dvipsnames]{xcolor}
\usepackage{hyperref}
\hypersetup{
    colorlinks,
    linkcolor={blue!50!black},
    citecolor={blue!50!black},
    urlcolor={blue!80!black}
}

\begin{document}

   \title{Sun-like stars show lowered X-ray activity together with over-rotation at beyond-solar ages}
   \titlerunning{Old Suns paper}

   \author{K.~Poppenhaeger\thanks{E-mail: kpoppenhaeger@aip.de}
          \inst{1, 2}
          \and
          N.~Ili\'c\inst{1}
          \and
          J.~van~Saders\inst{3}
          \and
          J.~D. Alvarado-G{\'o}mez\inst{1}
          \and
          F.~Anders\inst{4, 5, 6}
          \and
          M.~M.~Jardine\inst{7}
          \and
          A.~Warmuth\inst{1}
          \and
          E.~M.~Amazo-G{\'o}mez\inst{1}
          \and
          S.~A.~Barnes\inst{1}
        }

   \institute{Leibniz Institute for Astrophysics Potsdam (AIP), An der Sternwarte 16, 14482 Potsdam, Germany
        \and
            Institute for Physics and Astronomy, University of Potsdam, Karl-Liebknecht-Str. 24/25, 14476 Potsdam-Golm, Germany
        \and
            Institute for Astronomy, University of Hawai‘i, 2680 Woodlawn Drive, Honolulu, HI 96822, USA
        \and
            Departament de Física Quàntica i Astrofísica (FQA), Universitat de Barcelona (UB), Barcelona, Spain
        \and
            Institut de Ciències del Cosmos (ICCUB), Universitat de Barcelona (UB), Barcelona, Spain
        \and
            Institut d’Estudis Espacials de Catalunya (IEEC), Barcelona, Spain
        \and
            School of Physics and Astronomy, University of St Andrews, North Haugh, St Andrews, Fife KY16 YSS, UK
             }

   \date{Accepted Nov 02 2026. Received Oct 30 2026; in original form June 05 2026}

  \abstract
  {The rotational spin-down of cool main-sequence stars governs their long-term magnetic evolution and can even influence the survival of the atmospheres of planets orbiting them. In recent years, substructure in the age-rotation relationship of cool stars has been identified along both the stellar mass and age axes, showing deviations from the historical Skumanich law. A weakened magnetic braking regime and potentially a collapse of the magnetic dynamo have been proposed as reasons for this deviation.} {Our aim is to investigate the magnetic activity as a function of stellar rotation for sun-like stars older than the Sun. Since magnetic activity is expected to be the physical root cause of the stellar spin-down in this mass regime, namely via the magnetic braking process mediated through stellar winds, changes in spin-down should correspond to unusual activity behaviour.} {We selected a sample of old sun-like stars in the solar neighborhood with asteroseismic age estimates and measured rotation periods, which sit in the suspected weakened magnetic braking regime. We present an analysis of their magnetic activity using deep X-ray observations from XMM-Newton. We determine X-ray luminosities and infer coronal temperatures from the data and put them into context with the observed rotational behaviour.} {We observe that sample stars less massive than the Sun (late-G and early-K) agree well with the established rotation-activity relationship, while early-G and late-F stars with masses above solar tend to be unusually X-ray dim for their rotation period. The X-ray dim part of the sample contains the stars that are rotating faster than same-colour stars in the solar-age open cluster M67, despite being older than M67. Those stars are characterized by either low metallicity ([Fe/H]$\approx -0.2$\,dex compared to solar) or by a mix of low coronal temperatures ($< 1.5$\,million K) and very low X-ray surface fluxes ($\lesssim 10^4$\,erg\,s$^{-1}$\,cm$^{-2}$).} {We conclude that there is sufficient dynamo activity for coronal emission in the sample stars, since we successfully detect the majority of them in our X-ray observations. We argue that stars with extremely low coronal temperature and/or low X-ray flux through the stellar surface may produce an inefficient spin-down through hampered angular momentum shedding. If this is indeed the cause for their over-rotation, the onset age of this effect would need to be at an age of approximately 3~Gyr. In contrast, stars that are metal-poor may be afflicted by less efficient convection and shorter convective turnover times than stars of the same mass with solar metallicity, making them effectively behave like hotter stars in their spin-down behaviour, which provides a different explanation for their faster rotation. However, the behaviour of the part of our sample with solar or mildly super-solar metallicity is at odds with studies of stars in the open cluster M67 with a cluster age of $\sim4$~Gyr, which do not show a faster-rotating population of sun-like stars. To understand the rotational evolution of our Sun and solar analogue stars, the discrepancies to old open clusters need to be studied further.}

   \keywords{  Stars: activity --- Stars: solar-type ---  Stars: rotation ---  X-rays: stars  ---   Stars: magnetic field       }

   \maketitle

\section{Introduction}

Magnetic dynamos are ubiquitous among cool stars like the Sun. They cause not only a large-scale magnetic field, but also highly localized and time-variable phenomena like starspots, flares and a heating process that creates a hot chromosphere and even hotter corona. In the case of cool stars, stellar winds are launched from the corona, which creates a magnetic braking mechanism: the ionized wind couples to the stellar magnetic field as it moves away from the star, and it carries away angular momentum when it finally decouples from the magnetic field \citep{Parker1965, Mestel1968}. The consequence of this magnetic braking process is a significant spin-down of cool stars on timescales of gigayears. Historically, this was first formulated by \cite{Skumanich1972}, who observed both rotational broadening of sun-like stars and their chromospheric Ca\,II~H and K line emission to decrease with stellar age (with ages stemming from open stellar clusters). The overall time-dependence of the stellar rotation was proposed there to be proportional to the square-root of the stellar age. 

\cite{Barnes2003} showed through investigations of several open stellar clusters that a spin-down relationship exists also for cool stars of lower and higher mass than the Sun, with a distinct mass-dependence of the spin-down. It has become common practice thereafter to describe stellar spin-down in mass-age-rotation space, with stars clustering around a curved surface, as described in \cite{Meibom2015}. In recent years, rotation period measurements based on light curves from space missions such as Kepler and TESS have become available not just for young clusters and field stars (see for example \citealt{McQuillan2014, Rebull2016, Douglas2017, Douglas2019, Reinhold2023}), but also increasingly for stars in older open clusters (see for example \citealt{Barnes2016, Agueros2018, Curtis2020, Gruner2023}). Those studies have revealed that there is significant substructure in the spin-down relationship that deviates from the Skumanich-type square-root relation. Specifically, a possible delay in spin-down was identified for stars in the K dwarf regime around ages of 2-4~Gyr \citep{Curtis2020, Gordon2021, Gruner2023}, and a possible over-rotation for stars of spectral type G and late F at ages older than the solar age has been demonstrated \citep{vanSaders2016}. The former phenomenon has been mainly dubbed ``stalled spin-down'' and the latter ``weakened magnetic braking'' in discussions in the literature \citep{vanSaders2016, vanSaders2019, Curtis2020, Hall2021, Masuda2022, Saunders2024, Metcalfe2025}. While some works have started to use those terms interchangeably, they actually refer to different proposed origins of the over-rotation: stalled spin-down refers to an explanation emphasizing the angular momentum redistribution inside the star  (i.e.\ \citealt{Curtis2020}), and weakened magnetic braking refers to an interpretation focusing on the reduction in the angular momentum loss through the stellar wind (i.e.\ \citealt{vanSaders2016}).

Spin-down due to magnetic braking is physically governed by the magnetic activity of the star. Fast stellar rotation of cool stars causes high magnetic activity through the magnetic dynamo, which in turn is expected to cause strong spin-down due to efficient expulsion of a stellar wind. This gives rise to the well-studied rotation-activity relationship \citep{Noyes1984, Pizzolato2003, Wright2011, Reiners2022} and in combination with stellar ages to the age-rotation-activity relationship \citep{Guedel1997, Pace2004, Kiraga2007, Mamajek2008}. While the age-rotation-activity relationship seems rather straight-forward for moderate stellar ages, there are aspects that are not fully understood yet at young stellar ages and fast rotation periods. In terms of rotation-activity, fast-rotating stars display a saturation regime in magnetic activity \citep{Pizzolato2003, Wright2011}, where fast versus extremely fast rotation yields the same observed activity level; even a super-saturation regime for the fastest rotators has been proposed \citep{James2000MNRAS.318.1217J, Jardine2004A&A...414L...5J, Jeffries2011MNRAS.411.2099J}, although conclusive observational confirmation remains difficult due to the significant activity scatter in that regime. On the age-rotation side, young clusters display a sub-population of stars that have not converged yet onto the spin-down track, but seem to be ``stuck'' in fast rotation for a while; this behaviour lasts longer for stars with lower mass \citep{Barnes2003}. Hypotheses that have been brought forward to explain this behaviour centre around geometric effects of the stellar magnetic field that may affect the efficiency of angular momentum shedding through the wind \citep{Wright2011, Garraffo2016, Garraffo2018, Chebly2023}. However, analysis of Zeeman Doppler Imaging maps for cool stars points to dipolar components of the stellar magnetic field being the main providers of spin-down torques instead, even for younger stars \citep{See2019}.

In this work, we focus on the reported deviation from expected spin-down at spectral types G and late F and stellar ages at or beyond the solar age. Our aim is to investigate whether the magnetic X-ray activity of stars in this regime is unusual, indicating a different regime of activity-induced stellar winds and associated angular momentum loss, which in turn may provide an explanation for the stellar rotational behaviour. Previous studies showed that the observed X-ray activity of cool stars with ages beyond a gigayear decays steeply with increasing stellar age \citep{Booth2017}; modelling of stellar winds for sun-like stars of various ages and activity levels demonstrated that lower coronal temperatures, associated with lower magnetic activity overall, lead to a strongly reduced angular momentum loss of the star due to lower mass loss rates and stellar wind velocities at stellar ages beyond two gigayears \citep{Fionnagain2018}. Recently, a weakening of the large scale dipole field of old sun-like stars has been brought forward as an explanation of their faster-than-expected rotation, supported by some restrictive upper limits on the large scale magnetic field strength from spectropolarimetric observations (see for example \citealt{Metcalfe2024, Metcalfe2025}).  

A weakened or broken dynamo for such stars would make further spin-down impossible due to a nonexistent or severely suppressed corona and stellar wind, which should lead to observable consequences in the stellar X-ray emission. However, studying low magnetic activity in stars is pushing the flux limits of what current X-ray observatories can achieve. Detecting and characterizing X-ray emission from stars with a potentially weak dynamo even out to 100\,pc distance from the Sun requires very deep exposures with a significant observing time investment.

Nevertheless, X-ray observations are a diagnostically sensitive tool to study the magnetic activity of stars; unlike chromospheric activity indicators, stellar X-ray emission does not have a significant basal (non-activity-related) component. Furthermore, X-ray observations contain information about the temperature of the stellar corona, with higher coronal temperatures causing harder X-ray spectra. Coronal temperatures are different for solar surface features such as coronal holes and active regions, and result in different velocities of the solar wind being launched from those regions \citep{Finley2019, Verscharen2021}. For the stellar case, also the average stellar coronal temperatures have been investigated in terms of their ability to alter angular momentum loss \citep{Pantolmos2017, Fionnagain2018, Levesque2025}.

It is important to note that stars with low magnetic activity typically display low coronal temperatures and therefore very soft X-ray spectra. Main coronal emission lines like OVIII or the OVII triplet have their peak formation temperatures at ca.\ 3~MK and 2~MK, respectively. For stars with cooler coronae, their emission in those lines can be very weak and the bulk of the stellar emission is in fact seen at energies below 500~eV, such as for 51~Pegasi (\citealt{Poppenhaeger2009}, see also \citealt{Metcalfe2024}). It is therefore imperative to study the extremely soft X-ray emission of stars with suspected low activity, for which XMM-Newton is currently the most sensitive X-ray observatory in operation. In this work we describe the observations and the data analysis steps geared towards characterising such low-activity coronae in sections~\ref{sec:obs} and \ref{sec:methods}, report our results regarding the age-rotation-activity properties of the sample stars in section~\ref{sec:results}, and discuss potential explanations for the behaviour seen in section~\ref{sec:disc}. We end with our conclusions in section~\ref{sec:con} and display additional information and plots in the Appendix.

\section{Observations}\label{sec:obs}

XMM-Newton is an X-ray and optical-ultraviolet space observatory that carries several instruments collecting data from targets in parallel \citep{Jansen2001}. The most relevant instrument for the research presented here is the European Photon Imaging Camera (EPIC), consisting of three cameras, namely the MOS1 and MOS2 camera (using Metal Oxide Semi-conductor CCD arrays) and the PN camera (using pn-CCDs) \citep{EPIC_mos,EPIC_pn}. All three cameras are typically used in the energy band of 0.2-12~keV. However, they do have non-zero sensitivity at lower energies, with MOS1 and MOS2 being sensitive down to energies of 75~eV (albeit with low quantum efficiency at those soft energies) and PN down to energies of 150~eV. PN suffers from additional instrumental effects below 200~eV, most prominently detector noise patterns and spurious X-ray events caused by proton impacts from the solar wind, which require careful analysis (for an overview, see \citealt{Dennerl2004}). 

In addition, XMM-Newton carries two high-resolution Reflection Grating Spectrometers (RGS) for soft X-rays as well as an Optical Monitor (OM) \citep{XMM_rgs, XMM_om}. Data from those instruments is not analysed here, since the targets stars are too X-ray faint for any useful spectra from RGS, and the OM images were mostly saturated.

The core of the data analyzed here stems from an XMM-Newton Large Program (program ID 09211, PI Poppenhaeger), which collected deep exposures of six target stars with the EPIC instrument. In addition, several data sets from the XMM-Newton archive were analyzed, covering some of the same stars in shorter exposures and other stars with similar properties. Relevant information about the analyzed observations is given in the Appendix in Table~\ref{tab:obs}.

\section{Methodology}\label{sec:methods}

The aim of this work is to analyze the magnetic X-ray activity of old sun-like stars, and to determine whether they display anomalous age-rotation-activity behaviour. 

\subsection{Sample selection}\label{sec:methodology:sample}

To define our sample, we selected stars with existing asteroseismic age estimates and determined rotation periods that are located within up to 100\,pc distance from the Sun, in order to make an X-ray detection feasible even at low activity levels. Rotation periods detected through optical light curve modulation may bias such a sample, since the Sun itself would typically not produce a readily detectable rotation signature in light curves at data quality levels of current space-based photometric telescopes \citep{Shapiro2020, Amazo-Gomez2020a, Amazo-Gomez2020b}; therefore, old sun-like stars with a detected surface rotation signal may be unusual in their magnetic activity. However, rotation periods can also be detected through asteroseismic analysis due to rotational splitting of asteroseismic oscillation frequencies, independently from the presence of spots on the stellar surface \citep[e.g.][]{Ballot2006,Benomar2015, Davies2015}. 

We therefore used such a sample as our starting point, namely the sample presented by \cite{Hall2021}, who derived ages and rotation periods asteroseismically for a number of  main-sequence and evolved stars. We filtered their sample to main-sequence stars and stellar distances up to 100\,pc, stellar masses between 0.8 and 1.1\,$M_\odot$, asteroseismic ages of 5\,Gyr or older, and data quality flags of 0 or 1. This yielded a remaining sample of 10 stars, out of which 9 have been observed in X-rays with \textit{XMM-Newton} and which are further analyzed here. We manually added the star 51~Peg to the sample, which is another nearby star that has been studied asteroseismically in detail \citep{Metcalfe2024} and has well-determined low-level X-ray activity \citep{Poppenhaeger2009, Metcalfe2024}. We point out that our resulting sample is quite modest in size, and this is a fundamental limitation stemming from the accessible volume for X-ray observations of such stars. First, the flux limits of current X-ray observatories restrict at which distances stars at or below sun-like X-ray luminosity can still be detected. Second, stars with solar or sub-solar magnetic activity are expected to be dominantly soft X-ray emitters like the Sun itself. Since the interstellar medium absorbs soft X-ray photons more strongly than harder ones, very soft X-ray emitters at distances significantly larger than 100\,pc will be challenging to access even for future X-ray observatories with higher flux sensitivity overall.

We cross-matched those stars with the {\it Gaia} DR3 tables \verb|gaia_source| and \verb|astrophysical_parameters| \citep{GaiaDR3} to obtain additional stellar parameters of interest such as distances, {\it Gaia} colors, and bolometric luminosities. We also queried reddening values (\verb|ebpminrp_gspphot|) from {\it Gaia}, 
in order to be consistent with cluster rotation studies that use de-reddened {\it Gaia} $BP-RP$ colors; they consistently turned out to be very small ($\leq 0.054$\,mag), as expected for stars within 100\,pc around the Sun. We additionally obtained $V-K$ colors through SIMBAD, for comparisons with the well-studied large stellar sample from \citet[][W11 from here on]{Wright2011}.

We note that the stellar Rossby number instead of rotation period is often used in rotation-activity analyses, since it can in principle normalize across the different dynamo efficiencies of stars of different mass. However, this requires an estimate of the convective turnover time $\tau$, which is not a directly observable quantity. W11 use an empirical estimate for $\tau$, derived from minimizing the scatter in the observed activity-rotation characteristics of their sample for different stellar colour bins. Similar empirical approaches, such as \cite{Bonanno2025}, additionally include stellar metallicities in their estimates. Modelling-based estimates of $\tau$ take into account stellar surface gravities, effective temperatures and metallicities (see for example \citealt{vanSaders2013, Gossage2021}). The various estimates of $\tau$ can differ significantly, especially in the realm of F stars with shallow convection zones.

We therefore opted for an alternative approach presented by \cite{Reiners2014} that does not use the Rossby number, but uses the stellar rotation period alone. \cite{Reiners2014} demonstrated that in the so-called unsaturated regime, i.e.\ for stars that are not very fast rotators, the rotation-activity relationship can be expressed solely in terms of the stellar rotation period $P_{\mathrm{rot}}$ and X-ray luminosity $L_\mathrm{X}$. Not only does this approach have the immediate advantage of using only directly observable quantities, it also turns out that the observed relationship between $P_{\mathrm{rot}}$ and $L_\mathrm{X}$ contains actually less scatter than the more commonly used relationship between Rossby number and fractional X-ray luminosity $L_\mathrm{X}/L_{\mathrm{bol}}$. We therefore use stellar rotation and X-ray luminosity as our main parameters in this work, while we show our results in Rossby number and $L_\mathrm{X}/L_{\mathrm{bol}}$ space in the Appendix for comparability to other works.

\subsection{Two stars with questionable binary status: KIC~7871531 and KIC~9025370}

It is important for the interpretation of the X-ray measurements to ascertain that the sample stars are truly single stars. Close binary systems will have different rotational evolution due to tidal locking, and recent studies of stellar rotation in young associations, open clusters, as well as field stars found that out of the fast-rotation outliers, particularly the ones with rotation periods of two days or less, almost all can be confirmed to be close binaries \citep{2016ApJ...822...47D, 2019ApJ...871..174S, 2022AJ....164..137K, 2023AJ....165..182K}.
Also wide binary systems need to be treated with care: While many of the relevant stellar parameters are derived from asteroseismology and can in principle be disentangled also for multiple stellar components that are unresolved, this is not possible for the X-ray data analysis due to its limited spatial resolution. Any X-ray activity measurement for non-single stars would therefore be contaminated by the stellar companions and not be a reliable representation of the activity state of the sun-like star of interest.

Out of the sample stars, 16~Cyg~A is known to host an M~dwarf companion at ca.\ $5^{\prime\prime}$ separation, which is unresolved in \textit{XMM-Newton}. We demonstrate in Appendix~\ref{appendix:16cygc} based on \textit{Chandra} data with higher spatial resolution that this M dwarf is so X-ray faint that it is of no concern to our analysis here. We can therefore treat the X-ray flux of 16~Cyg~A measured by \textit{XMM-Newton} as a reliable measurement of this G star's activity level.

Upon re-investigating the stellar properties of the remaining sample stars with {\it Gaia} DR3 data \citep{GaiaDR3}, we found that two additional stars display conspicuous behaviour. Specifically, KIC~9025370 is flagged as a non-single star in {\it Gaia} DR3 and was identified to be a double-lined binary by \cite{Thygesen2015}. KIC~7871531 has a very high {\it Gaia} DR3 \texttt{ruwe} (interpretable as a reduced $\chi^2$ of an astrometric single-star solution) value of 9.7, making its astrometric solution, distance and therefore most stellar parameters highly uncertain due to a higher probability of an unresolved stellar companion.

For completeness, we report the X-ray luminosities of those two stars in Table~\ref{tab:results} and show their X-ray images in in the Appendix, Fig.~\ref{fig:discarded}, but do not consider them further in the age-rotation-activity analysis of this work, since they are unlikely to be single old sun-like stars.

\subsection{X-ray data reduction}

\subsubsection{Choice of X-ray energy bands}

Ultrasoft X-ray photons, i.e.\ with energies below 200~eV, can make up substantial parts of the total X-ray emission for sun-like stars with low activity, comparable to the Sun at its activity minimum or in solar coronal holes (see for example \citealt{Poppenhaeger2009}). We therefore take particular care in our data reduction to measure the count rates observed by XMM-Newton's EPIC instrument also at energies below 200~eV, in addition to the energy bands traditionally used for stellar coronae. The traditionally used band is the 0.2-2~keV band or, for higher activity stars, harder bands above 2~keV as well.

We first perform a basic data reduction with the XMM-Newton Science Analysis Software (SAS) version 21.0.0 as specified in the SAS user guide\footnote{\url{https://xmm-tools.cosmos.esa.int/external/xmm_user_support/documentation/sas_usg/USG/}} to extract photon event files for the MOS1, MOS1, and PN cameras. We do not filter out events with energies below 200~eV, as we want to use those ultrasoft photons later in the scientific analysis. We perform a barycentric correction on the data using the SAS-implemented task \texttt{barycen}, which uses the JPL ephemeris data set and takes into account the space observatory's motion. We then proceed with the source detection, for which we use the SAS meta-task \verb|edetect_stack|.

\subsubsection{X-ray source detections and upper limits}

The \verb|edetect_stack| task is the most sophisticated source detection procedure implemented in SAS and is described in detail in \cite{Traulsen2019}. In short, it combines several observations of a target and performs a series of detection and background fitting steps in user-defined energy bands, ending in a point spread function (PSF) fit for the detected sources, which then yields detection likelihoods, X-ray count rates and associated count rate uncertainties for the sources.

\begin{figure}
 \includegraphics[width=0.5\textwidth]{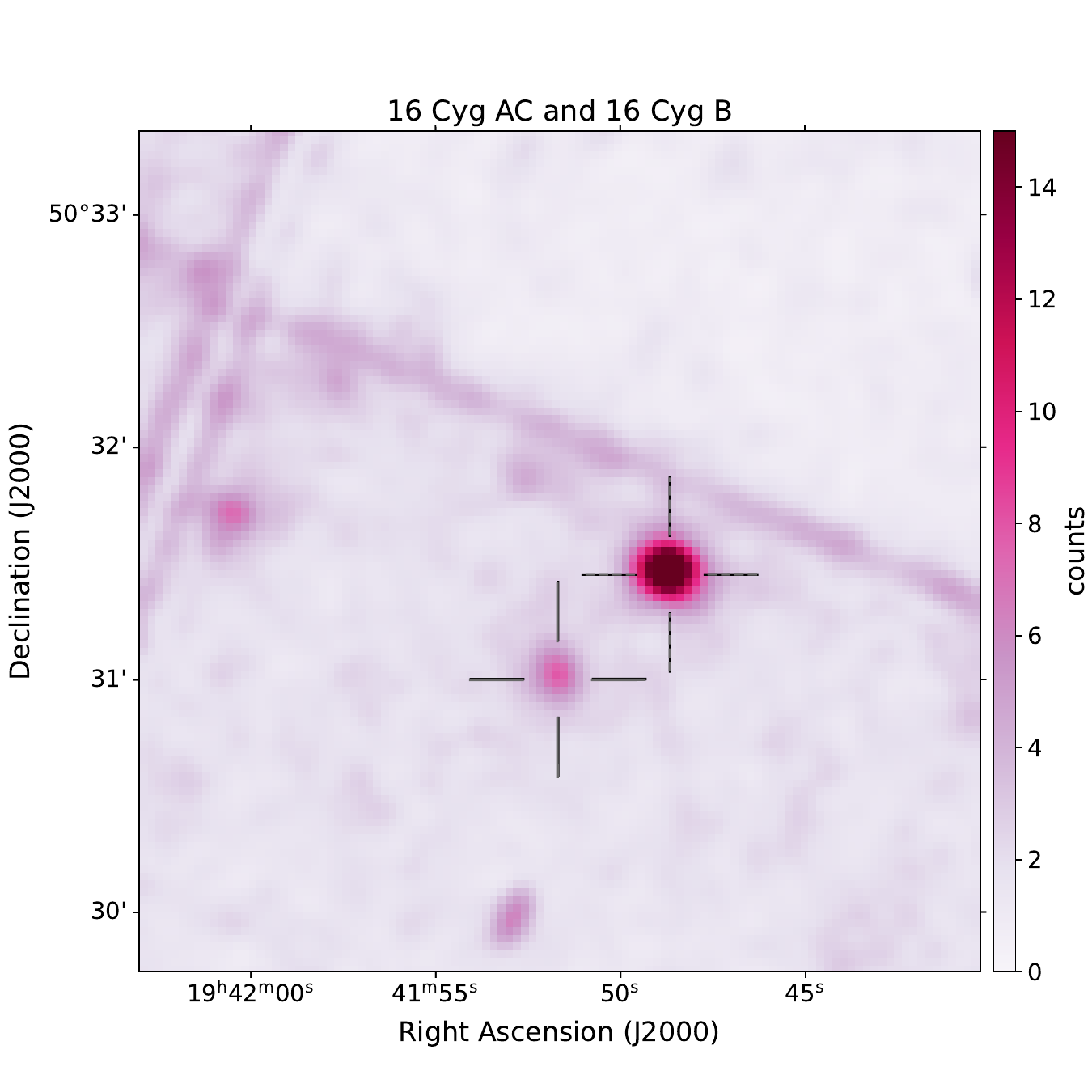} 
 \caption{The stars 16 Cyg A and B as seen in the X-ray data. The dashed-line reticules indicate the expected position of the two stars from {\it Gaia} at the time of the X-ray observation. The shown X-ray data are the combined MOS1, MOS2 and pn photons for the 75-2000~eV band. The X-ray brighter source is 16~Cyg~A (together with its spatially unresolved, low-mass C component; \citealt{Hauser1999}), the X-ray dimmer source is 16~Cyg~B. The unresolved C component is undetected in spatially resolved X-ray observations with \textit{Chandra} and is therefore unproblematic for our analysis (see Appendix~\ref{appendix:16cygc}).}
 \label{fig:16cygxmm}
\end{figure}

\begin{table*}
 \small \caption{Selected properties of the target stars. X-ray luminosities and fluxes are given for the 0.1-2.4 keV energy band. See text for details. \label{tab:results}}
\begin{tabular}{l r r r r r r r r r}
\hline\hline
name & index & (BP-RP)$_0$ & age & $P_{rot}$ & hardness ratio & $T_{corona}$ & $\log L_{X}$ & $\log(L_{X}/L_{bol})$ & $F_{X,\,surf}$ \\
 &  & [mag] & [Gyr] & [d] &  & [MK] & [erg\,s$^{-1}$] &  & [erg\,s$^{-1}$\,cm$^{-2}$] \\
 & & (Gaia) & (H21) & (H21) & (this work) & (this work) & (this work) & (this work) & (this work) \\ \hline
KIC 6106415 & 1 & 0.72 & $5.0^{+1.3}_{-1.1}$ & $16.0^{+0.7}_{-0.8}$ & $0.13 \pm 0.3$ & $3.0^{+ 0.67}_{-0.9}$ & $27.01$ & $-6.85$ & $10053$ \\
KIC 6116048 & 2 & 0.73 & $9.6^{+2.2}_{-1.9}$ & $17.9^{+0.8}_{-1.2}$ & -- & -- & $<26.93$ & $<-6.92$ & $<8507$ \\
KIC 9955598 & 3 & 0.92 & $6.3^{+1.9}_{-1.8}$ & $31.4^{+9.1}_{-6.4}$ & $0.08 \pm 0.3$ & $2.68^{+ 1.0}_{-0.89}$ & $27.21$ & $-6.18$ & $31706$ \\
KIC 10963065 & 4 & 0.7 & $7.2^{+1.9}_{-1.6}$ & $11.5^{+1.3}_{-1.0}$ & -- & -- & $<27.49$ & $<-6.38$ & $<31345$ \\
KIC 11772920 & 5 & 0.92 & $10.7^{+2.7}_{-3.0}$ & $35.1^{+5.3}_{-4.7}$ & -- & -- & $<27.36$ & $<-5.98$ & $<50286$ \\
16 Cyg A & 6 & 0.81 & $6.7^{+0.8}_{-0.8}$ & $20.5^{+2.0}_{-1.1}$ & $-0.18 \pm 0.0$ & $1.86^{+ 0.1}_{-0.11}$ & $27.62$ & $-6.16$ & $40722$ \\
16 Cyg B & 7 & 0.83 & $7.4^{+0.9}_{-0.9}$ & $21.2^{+1.8}_{-1.5}$ & $-0.47 \pm 0.1$ & $1.39^{+ 0.1}_{-0.1}$ & $27.17$ & $-6.5$ & $18093$ \\
51 Peg & 8 & 0.83 & $4.8^{+0.7}_{-0.4}$ & $21.9^{+0.4}_{-0.4}$ & $-0.77 \pm 0.2$ & $1.05^{+ 0.12}_{-0.02}$ & $26.57$ & $-7.15$ & $4151$ \\
\multicolumn{4}{l}{\textit{excluded from sample:}} & & & & \\
KIC 7871531 & 9 & 0.94 & $10.0^{+1.9}_{-1.8}$ & $33.1^{+4.1}_{-4.1}$ & $-0.15 \pm 0.1$ & $2.01^{+ 0.28}_{-0.35}$ & $27.96$ & $-5.48$ & $139038$ \\
KIC 9025370 & 10 & 0.84 & $6.5^{+1.3}_{-1.1}$ & $24.7^{+3.7}_{-4.7}$ & $-0.47 \pm 0.1$ & $1.41^{+ 0.16}_{-0.17}$ & $27.98$ & $-5.82$ & $84056$ \\
\hline
\end{tabular}
\end{table*}

Since we expect our target stars to display faint and very soft X-ray emission, we ran \verb|edetect_stack| first on the energy band 75-2000~eV, i.e.\ encompassing the lowest energies that are still being detected by the MOS cameras. The flux calibration of photons with energies below 200~eV is less accurate than for the 200-2000~eV photons. However, since out first goal was to ascertain whether the target stars were detected in X-rays at all, we took all soft photons into account. Since our targets all have well-defined positions in the sky from optical catalogs such as {\it Gaia} \citep{GaiaDR3}, we opted for a relatively lax detection threshold in the initial detection steps in \verb|edetect_stack| with a minimum detection likelihood $\geq 4$, and for the PSF fitting process we assumed no intrinsic source extension (since our stars are point sources) and a minimum detection likelihood $\geq 6$.

The lists of significantly detected sources from \verb|edetect_stack| were checked visually against the respective extracted X-ray images and the expected source positions at the epoch of the observations, which we derived from Gaia DR3 coordinates and proper motions \citep{GaiaDR3}. One example of the extracted X-ray images is shown in Fig.~\ref{fig:16cygxmm}, the remaining ones are shown in the Appendix in Fig.~\ref{fig:detections} and \ref{fig:non-detections}. We found that five of the eight sample stars were significantly detected at positions less than $3^{\prime\prime}$ separation from the expected {\it Gaia} position; those stars are 51~Peg, 16~Cyg~A, 16~Cyg~B, KIC~6106415, and KIC~9955598. This is in line with the median positional accuracy found in the 4XMM-DR12 catalog of $\leq 1.6^{\prime\prime}$ \citep{4XMM_Webb2020}, and we regard these stars to have bona fide X-ray detections in our observations. Observations of two stars only yielded significant detections with a separation of more than $9^{\prime\prime}$ from the expected position (KIC~10963065 and KIC~11772920), and we therefore do not regard these two stars to have X-ray detections. For the star KIC~6116048, our analysis yielded a nominal detection close to the expected position, but visual inspection showed an excess photon pattern that is not well explained by the XMM-Newton PSF for a point source (see Fig.~\ref{fig:non-detections}). There are also no other sources within a $10^{\prime\prime}$ radius around the star's position given by {\it Gaia} DR3, which could explain the apparent side-lobes seen in the X-ray image. We therefore also regard this star to not have an X-ray detection in our observations. 

To derive upper limits on the count rates for those undetected stars, we used \verb|edetect_stack| to fit a point source at the fixed optical position of the star in addition to the significantly detected sources nearby. As shown by \cite{Quintin2024}, fitting a null flux in this manner and taking its $3\sigma$ uncertainty yields a reasonable estimate of the upper limit of the source's X-ray count rate. We applied this methodology to determine upper limits to EPIC count rates for KIC~10963065 and KIC~11772920; for KIC~6116048, which nominally has a detected count rate excess at the optical source position, we take that detected rate as the upper limit.

\subsubsection{X-ray fluxes}

To derive X-ray fluxes from the detected count rates, an underlying X-ray spectrum needs to be assumed. Stellar coronae are well described by an optically thin thermal plasma model, the major determining parameter being the coronal temperature. The coronal temperature components can usually be determined from X-ray spectra; however, since most of our sources are very faint, no meaningful X-ray spectrum can be extracted for almost all of our targets (except for 16 Cyg A and B, see Appendix~\ref{app:fluxes}). We therefore chose to determine a probable underlying spectral shape for each detected star from a measured hardness ratio of two energy bands. Specifically, we re-ran \verb|edetect_stack| for the energy band of 0.15-2.0~keV, which represents a sweet spot for our analysis since it allows us to collect a good number of photons below 0.2~keV, but does not include even softer energies which can be assumed to have poor instrumental calibration. We then split this band into a soft band of 0.15-0.6\,keV and a hard band of 0.6-2.0\,keV. The motivation for the choice of these bands and the procedure to determine X-ray fluxes for the sample stars are described in Appendix~\ref{app:fluxes}.

With the approximate coronal properties determined as described in Appendix~\ref{app:fluxes}, we transformed the detected count rates for each star in the 0.15-2.0\,keV energy band into fluxes in a variety of soft X-ray bands that are used in the literature, namely 0.1-2.4\,keV (the \textit{ROSAT} band), 0.2-2~\,keV (the soft \textit{XMM-Newton} band), and the custom 0.15-2\,keV band used here. 

For the stars that did not yield an X-ray detection, we assumed that their underlying spectrum is very soft, since a correlation between X-ray spectral hardness and stellar X-ray luminosity is well established in the literature \citep{Schmitt1997, Telleschi2005, Johnstone2015}. We assigned those undetected stars the same flux conversion factor that we inferred for our softest detected star, which was 51~Peg. We then used this conversion factor to transform the upper limits on EPIC count rates for the 0.15-2~keV energy band to an upper limit on the fluxes in the various output energy bands.

\subsection{Fundamental properties of the sample stars}

\begin{figure}
 \includegraphics[width=0.45\textwidth]{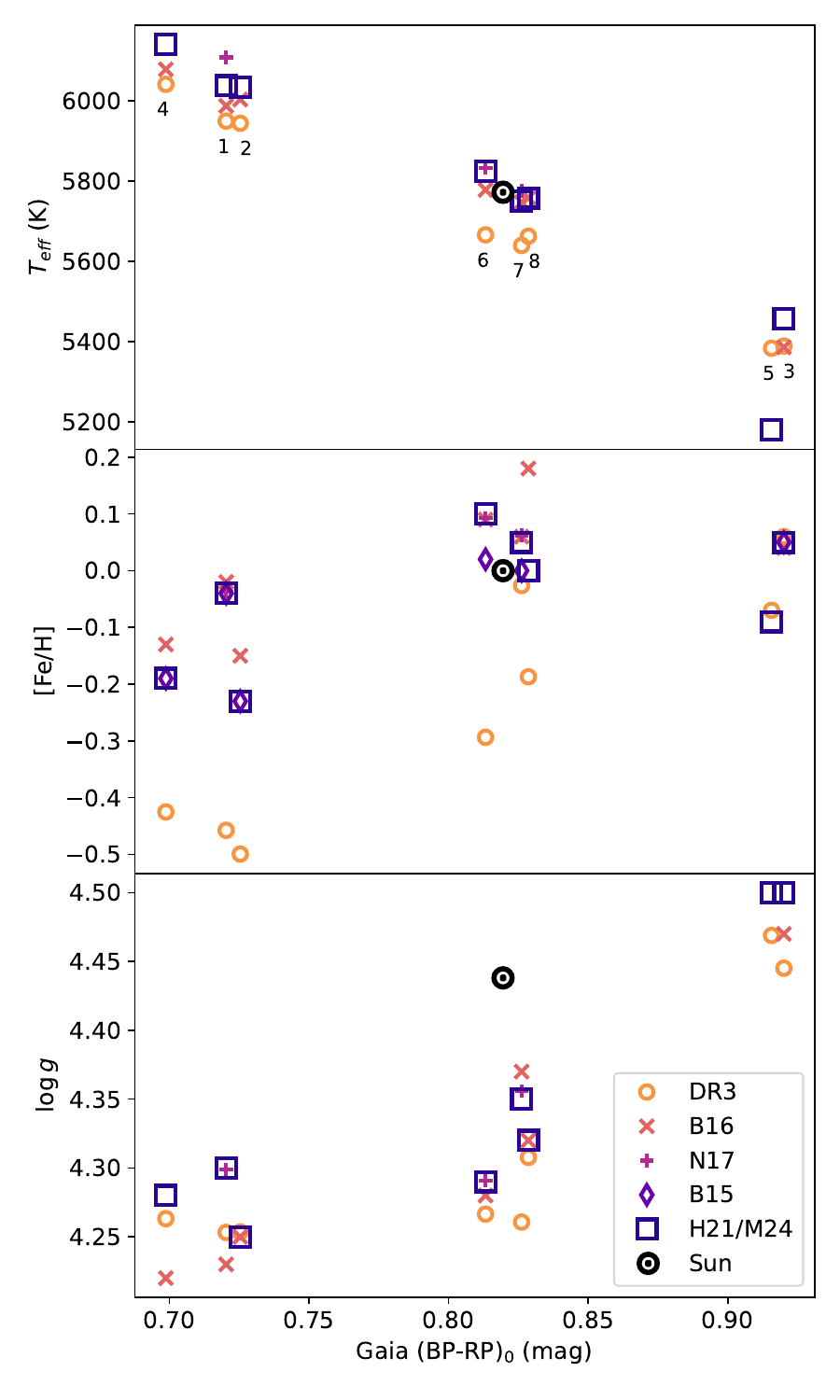}
 \caption{Comparison of fundamental stellar parameters of the sample stars from various sources (DR3-GSP = GSP values from \citealt{GaiaDR3}, B16 = \citealt{Brewer2016}, N17 = \citealt{Nissen2017}, B15 = \citealt{Buchhave2015}, H21 = \citealt{Hall2021}, M24 = \citealt{Metcalfe2024}). Shown are effective temperatures, metallicities and surface gravities. The numbers in the plot identify each star according to the ``index'' column in Table~\ref{tab:results}.}
 \label{fig:fundprops}
\end{figure}

To check whether the sample stars are unusual in their fundamental properties, we used stellar properties for the sample stars from \cite{Hall2021} as our baseline, with the exception of for 51~Peg, where we took those properties from \cite{Metcalfe2024}. To put these into context, we collected other determinations of their effective temperature, surface gravity and metallicity from the literature. Specifically, we consulted \cite{Nissen2017}, \cite{Brewer2016}, \cite{Buchhave2015} and the Gaia DR3 stellar properties derived from the Generalized Stellar Parametrizer GSP-Phot \citep{GaiaDR3, GaiaGSPPhot_Liu2012}; however, we will see shortly that the {\it Gaia} spectro-photometric metallicity estimates disagree quite strongly with the other metallicity determinations. Those sources will be abbreviated as H21/M24, N17, B16, B15 and DR3-GSP from here on. Furthermore, we added the Sun for comparison \citep{Prsa2016, Casagrande2018}. We display the various published values in Fig.~\ref{fig:fundprops}. 

From those comparisons, we can see the following:

a) The effective temperature determinations of the sample stars agree quite well (within $\sim100$ K) across all selected literature sources.

b) The metallicity determinations from high-resolution spectroscopy (N17, B15, B16) overall agree well with the values given by H21. The {\it Gaia} DR3-GSPPhot metallicities, however, sit at significantly lower values than the high-resolution spectroscopic ones for most stars, as also found in \citet[][see their Fig. 34]{Babusiaux2023}. We assume the high-resolution spectroscopic ones to be more reliable than the DR3 ones, and do not take into account the {\it Gaia} metallicities further in our interpretation. We also see that there are a few sample stars with significantly sub-solar metallicity around $-0.2$~dex, namely KIC~6116048 and KIC~10963065 (star 2 and 4, respectively). This can be relevant in the context of stellar rotational behaviour, because a lower metallicity would mean a shallower convection zone and shorter convective turnover timescale of the star, which in turn would make a star behave more ``F-star like'' than its colour would suggest.

c) The stars that are similar in colour to the Sun or bluer are slightly more evolved, as seen in the surface gravities. The surface gravities of those stars sit between the solar level and the beginning of the subgiant branch for the respective effective temperatures. This is not very surprising, since most of those stars have older age estimates than the solar age. This, however, does not alleviate the tension between the predicted rotation periods from the open clusters and the observed stellar rotation. Evolutionary models for stars in the solar mass regime show monotone increases in rotation period as the stars move onto the subgiant branch, see for example \cite{doNascimento2012}. Therefore, one would expect additional spin-down due to internal and radius changes of the star as it begins to evolve off the main sequence.

\begin{figure}
 \includegraphics[width=0.49\textwidth]{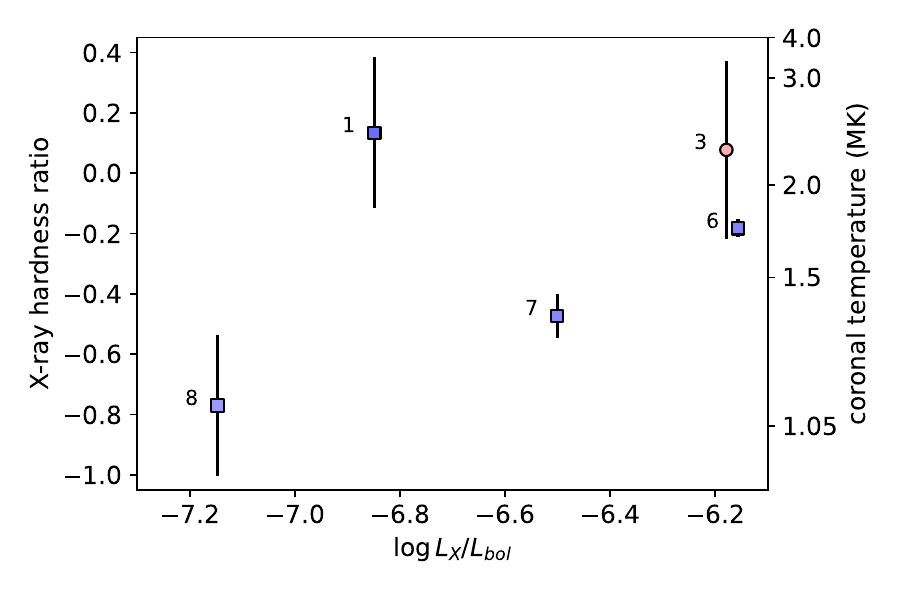}
  \caption{The observed XMM-EPIC X-ray hardness ratios of the successfully detected sample stars increase with increasing X-ray activity indicator $\log L_X/L_{bol}$ (albeit with large scatter), which is the standard behaviour seen for cool stars (see for example \citealt{Johnstone2015}). The hardness ratios used were 0.15-0.6 keV for the soft band and 0.6-2.0 keV for the hard band. The numbers are the same as the star index in Table~\ref{tab:results}, the colour coding of the individual stars is the same as the one chosen in Fig.~\ref{fig:hyperplane}; see text in section~\ref{sect:age-rot} for details. On the second y axis, the corresponding average coronal temperatures for those hardness ratios are shown.}
 \label{fig:hardnessratio_activity}
\end{figure}

\begin{figure*}[ht!]
\begin{centering}
 \includegraphics[width=0.9\textwidth]{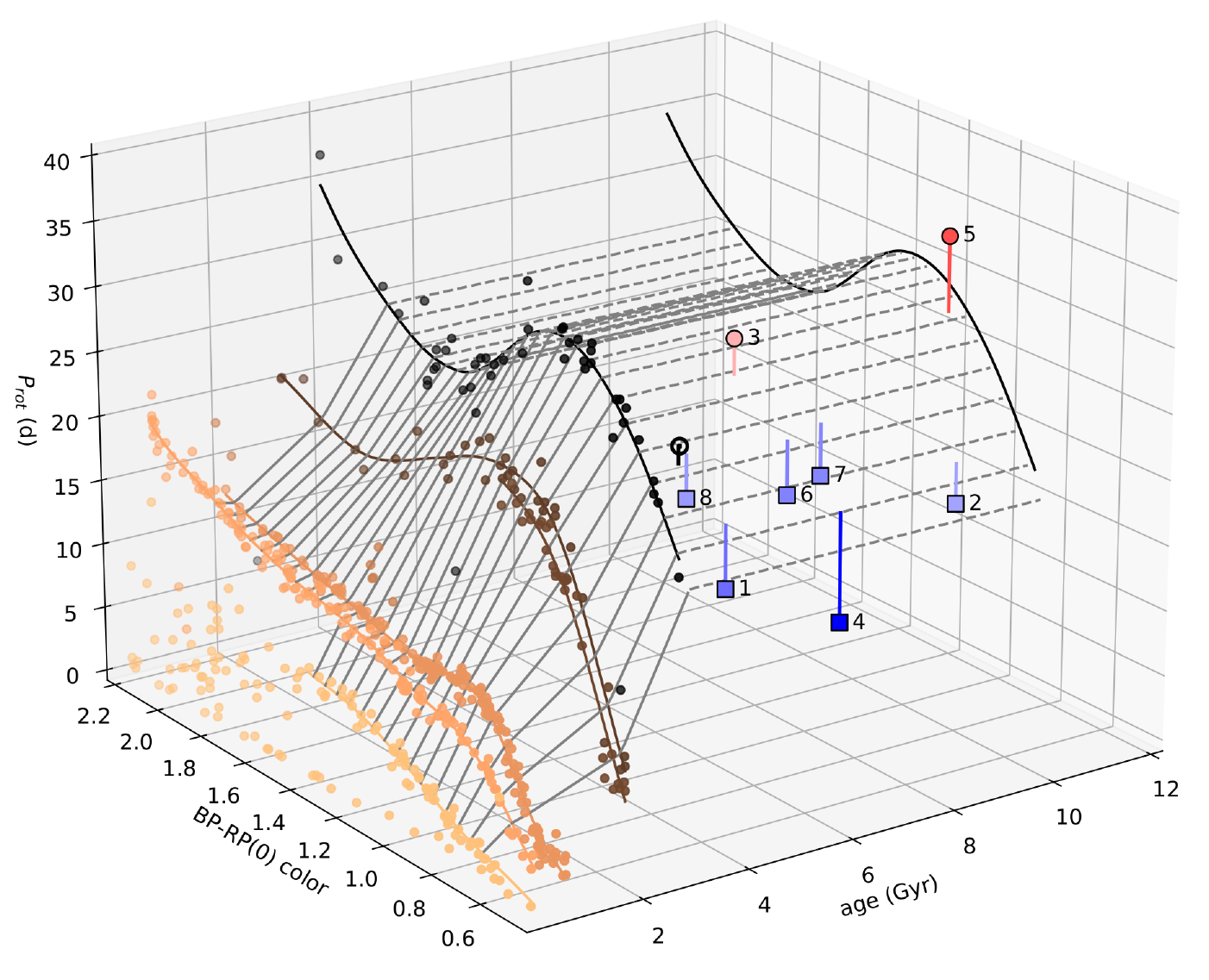}
  \caption{The sample stars (large circles and squares) are displayed in age-colour-rotation space, together with a series of stars from open stellar clusters (orange, brown and black small dots). The clusters shown are the Pleiades (0.12~Gyr), Praesepe (0.67~Gyr), NGC~6811 (1~Gyr), NGC 6819 (2.5~Gyr), Ruprecht~147 (2.7~Gyr) and M67 (4~Gyr), cluster data from \cite{Curtis2020} and \cite{Gruner2023}; solid grey lines are iso-colour lines. To guide the eye, we have plotted the colour-rotation sequence of M67 again at an age of 11~Gyrs, to make it easier to discern whether the sample stars sit below or above the plane indicated by the dashed iso-colour lines. We colour-code the sample stars by their deviation from the plane, with red circles sitting at longer rotation periods above the plane and blue squares at shorter rotation periods below the plane. The majority of the sample stars spin faster than stars of the same $(BP-RP)_0$ colour in M67.}
  \label{fig:hyperplane}
  
\end{centering}

\end{figure*}

We also checked whether the stars are unusual in terms of known exoplanets and their planetary parameters. This could in principle be relevant, since Star-Planet Interactions (SPI) may cause higher stellar activity and rotation rates \citep{Cuntz2000, Shkolnik2005, Shkolnik2008, Pont2009}. Tidal spin-up due to close-in planets has been identified \citep{Ilic2022}, and recently also high stellar flare rates, likely magnetically triggered by an exoplanet, have been demonstrated \citep{Ilin2025nature}. Our sample of stars contains four planet hosts: 16~Cyg~B, 51~Peg, KIC~9955598 (also known as Kepler-409), and KIC~10963065 (also known as Kepler-408). Two of those stars (16~Cyg~B, KIC~9955598) host known planets that orbit them in such a wide orbit ($P_{orb}>50$\,d, see \citealt{Cochran1997} and \citealt{Marcy2014} for the discovery papers) that SPI should not be relevant there. KIC~10963065's planet is in a short $2.47$\,d orbit, but is smaller in radius than Earth \citep{Marcy2014} and therefore expected to be of such low mass that again SPI should not be relevant to the stellar rotation. Only the seminal 51~Peg system hosts a known Hot Jupiter \citep{MayorQueloz1995}, and it is the only star in the sample where a relevant angular momentum transfer is at least thinkable (see \citealt{Ilic2024} for a discussion). At the same time, 51~Peg displays very low stellar activity, so if there is a tidal spin-up due to SPI, it apparently does not manifest itself as a simultaneous increase in stellar activity as seen in other systems with indications for SPI. Magnetic SPI would also typically be expected to increase the activity level for these types of stars, which is not what we will see in our following analysis. Therefore, all in all we conclude that SPI is not a relevant factor for the behaviour of our sample as a whole.

\section{Results}\label{sec:results}

\subsection{The magnetic activity of the target stars in the X-ray band}

Our remaining sample of eight stars yielded five X-ray detections and three upper limits. 
Using stellar distances from {\it Gaia} DR3, we transformed the derived X-ray fluxes into X-ray luminosities for the sample stars for a number of relevant energy bands. All of the sample stars display moderate to low X-ray luminosities of $L_X \lesssim 5\times 10^{27}$~erg/s in the commonly used 0.1-2.4~keV energy band. We list the stellar X-ray properties together with some relevant stellar properties in Table~\ref{tab:results}. Specifically, we report there the de-reddened stellar colour from {\it Gaia}, $(BP-RP)_0$, the stellar age and rotation period as estimated from asteroseismic analysis from \cite{Hall2021}, and several X-ray properties we derived in this work, namely the hardness ratio $HR=(H-S)/(H+S)$ (with $S$ being the net source count rate in the soft energy band of $0.15-0.6$\,keV and $H$ in the hard energy band of $0.6-2.0$\,keV), the mean coronal temperature inferred from the hardness ratio (see Appendix~\ref{appendix:16cygc}), and the X-ray luminosity, X-ray over bolometric luminosity, and the X-ray flux through the stellar photospheric surface area, all in the $0.1-2.4$\,keV energy band. The stellar radius used for determining the stellar surface area stems from {\it Gaia} DR3; specifically, we use the radius estimated with the "FLAME" algorithm (\verb|gaiadr3.astrophysical_parameters.radius_flame|; see \citealt{Creevey2023}). We also supply a table with additional information, including ancillary stellar parameters and X-ray luminosities in other commonly used energy bands, in Appendix Table~\ref{tab:allresults}.

We investigated whether the spectral properties of the stars correlate with their X-ray activity level. It is well-known for cool stars that stars with higher X-ray activity, parametrized either as $L_X$, $L_X/L_{bol}$ or $L_X/A_\ast$ (with $A_\ast$ being the photospheric surface area of the star), display harder X-ray spectra corresponding to higher coronal temperatures (see for example \citealt{Johnstone2015}). We show the X-ray hardness ratios of our detected stars, with 0.15-0.6\,keV being the soft band and 0.6-2.0\,keV being the hard band, versus the fractional X-ray activity $L_X/L_{bol}$ in Fig.~\ref{fig:hardnessratio_activity}. There is a noticeable trend for lower hardness ratios at lower activity, as known from other cool stars; however, there is some significant scatter as well. The star KIC~6106415 may be a mild outlier from the otherwise present trend, but this is difficult to ascertain with such a small sample size.

\subsection{Age-Rotation behaviour of the sample stars}\label{sect:age-rot}

\begin{figure}
\begin{centering}
 \includegraphics[width=0.5\textwidth]{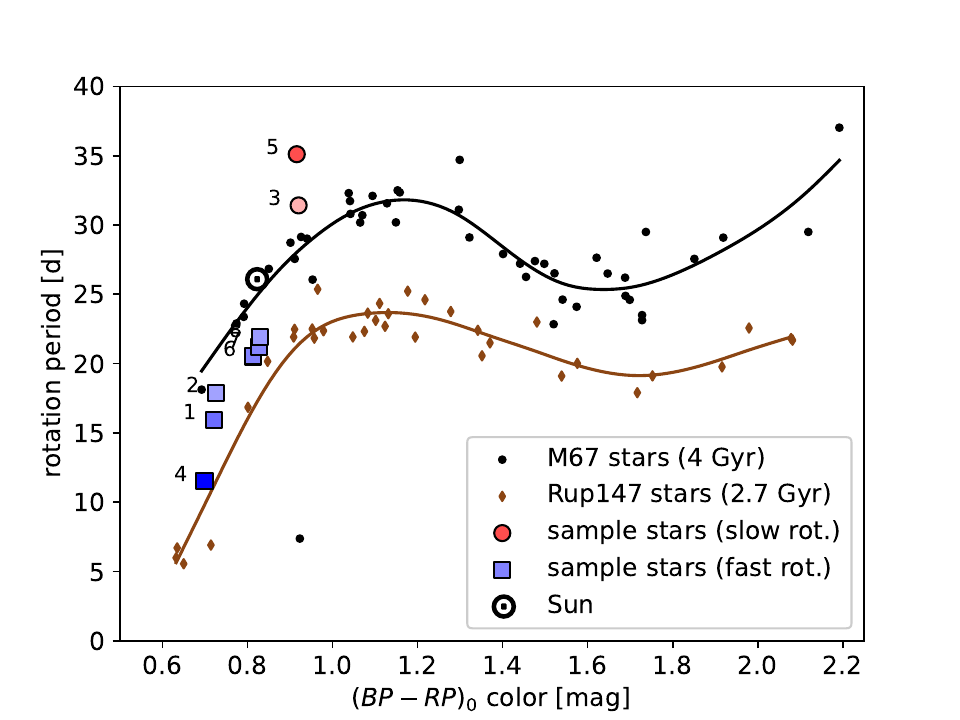}
  \caption{The sample stars are shown in colour-rotation space (with the age axis collapsed), together with the two clusters M67 and Ruprecht~147. Note that the colour axis is flipped compared to Fig.~\ref{fig:hyperplane}, to follow the conventional way of plotting the colour-rotation relationship. The discrepancy of the sample stars' rotation compared to M67 has a strong colour dependency (see Appendix Fig.~\ref{fig:periods_zoomed_with_uncertainties} for a zoom-in including uncertainties on periods where available.).}
 \label{fig:m67rotation}
 \end{centering}
\end{figure}

To place the sample stars into context, we compare them first to the age-rotation behaviour observed in cool stars using open clusters, and then to the rotation-activity relationship which uses both cluster and field stars.

Specifically, we take published rotation periods of cool stars in open stellar clusters from \cite{Curtis2020} and \cite{Gruner2023} and plot them as a function of cluster age and a parameter that scales with stellar mass in Fig.~\ref{fig:hyperplane}. Since stellar masses are not directly observable, but are model-dependent and may vary between different publications, we use the de-reddened Gaia $(BP-RP)_0$ colour instead of stellar mass. For the sample stars, de-reddening is not expected to be significant since they are all located within 100\,pc around the Sun, but for consistency we de-redden them via Gaia DR3 extinction values (\verb|ebpminrp_gspphot|), even though the de-reddening is very small for the sample stars.

It is well-studied that younger open clusters host a ``fast'' and a ``slow'' rotator stellar population \citep{Barnes2003}. An age-rotation relationship in the style of \citet{Skumanich1972} has only been established for the slow rotators; see for example \cite{Gruner2026}. To guide the eye, we therefore select each clusters' slow rotator members (excluding the M dwarfs which are not relevant for our analysis) and fit their rotation periods as a function of stellar colour with a smoothing spline. We display the cluster stars and the fitted curves in Fig.~\ref{fig:hyperplane}, together with connecting iso-colour lines to indicate the shape of the stellar plane in age-colour-rotation space, in a similar manner as presented by \cite{Meibom2015} for stars up to age 2.5~Gyr.

Our sample stars all have older asteroseismic age estimates than the age of the oldest cluster in the dataset, which is M67 with an age of approximately 4~Gyr \citep{Gruner2023, Reyes2025}\footnote{The exact age of M67 remains debated in the open-cluster literature, with the bulk of literature ages ranging from 3.5 to 4.5 Gyr (e.g. \citealt{Bossini2019, Li2025, Malhotra2026}), but also values between 2.5 and 5.2 Gyr have been given in some works (see \url{https://ucc.ar/_clusters/ngc2682/})}. Since the age-rotation relationship is not well constrained beyond that age \citep{Gruner2026}, we instead display a ''constant-M67-plane'' (indicated by dashed lines) that shows how the M67 stars would rotate if they did not spin down any more after 4~Gyrs, again to guide the eye.

The majority of the sample stars are located below this constant-M67-plane. This is not particularly surprising; both \cite{vanSaders2016} and \cite{Hall2021} identified those stars to be faster rotating than expected. However, in those works, the stellar rotational behaviour was compared to a Skumanich-type spin-down progression. In the meantime, detailed studies of stellar rotation in middle-aged open clusters have identified more substructure in the spin-down behaviour of cool stars \citep{Gruner2023, Curtis2020}. We therefore compare our sample stars to those empirically determined values, as shown in Fig.~\ref{fig:hyperplane}. We mention for completeness that stars around spectral type K (Gaia $(BP-RP)_0$ colours from 1.5 to 1.9 mag, approximately) display much milder spin-down than expected from the Skumanich relation at ages between 2 and 4 Gyr, as clearly visible in the middle-aged clusters M67 (black in Fig.~\ref{fig:hyperplane}) and Ruprecht~147 (brown in Fig.~\ref{fig:hyperplane}). However, by design our sample stars do not fall into the colour range of the plane where this additional substructure has been identified.

We colour-coded the sample stars by their deviation from the observed colour-rotation relationship in M67, with stars rotating faster than their counterparts in M67 being shown in blue and the ones rotating more slowly being shown in red. We will continue this colouring of the individual stars in all following plots of the analysis for easy comparisons.

We also show the sample stars in the context of the colour-rotation sequences of M67 and Ruprecht~147, having established ages of 4.0 and 2.7~Gyr, respectively \citep{Gruner2023, Curtis2020}, and the Sun in Fig.~\ref{fig:m67rotation}. The sample stars, despite having asteroseismic age estimates well above 4~Gyr, show an apparent colour-dependent deviation pattern from the rotation sequence seen in M67. The bluer, more massive stars rotate faster than expected, while the stars that are redder and less massive than the Sun display a spin-down relative to their M67 counterparts.

The less massive stars therefore display more of an expected behaviour -- being older than M67, they should experience some form of continued spin-down, even though their magnetic braking might be weakened. The more massive stars, however, display rotational behaviour that is \textit{unexpected under both Skumanich-braking and weakened magnetic braking}: if the sample stars and the stars in M67 follow the same braking law, then an older sample star should at most be rotating as fast as its younger M67 twin; this would be expected for the extreme case of no magnetic braking happening at all after the M67 age. If there is still residual magnetic braking, then the sample star should be rotating slower than its M67 twin. A situation like we have here, with older stars rotating faster than their younger twins, requires either a difference in magnetic braking laws that the sample stars and the M67 stars experience, or some form of spin-up or internal angular momentum redistribution for the older stars.  We discuss the implications of this surprising finding in more detail in section~\ref{disc:m67}.

\subsection{Rotation-Activity behaviour of the sample stars}

\begin{figure}
\begin{centering}
 \includegraphics[width=0.5\textwidth]{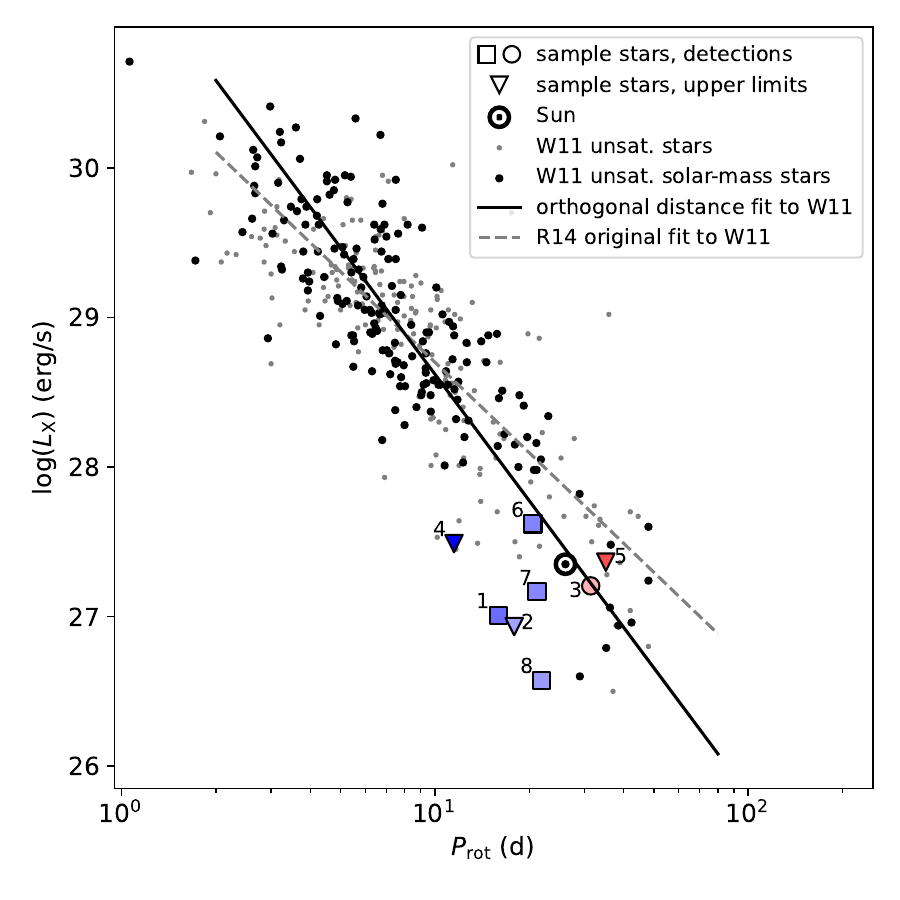}
  \caption{The X-ray luminosity $\log L_X$ of the target stars in the 0.1-2.4\,keV energy band is shown as a function of the stellar rotation period, together with the stars from the unsaturated regime of the W11 sample, similar to \cite{Reiners2014}, with solar-mass W11 stars chosen as having $0.8-1.1\, M_\odot$. The color-coding of our target stars is the same as in Fig.~\ref{fig:hyperplane}, squares again mean fast rotation, circles mean slow rotation; triangles denote X-ray upper limits. The original rotation-activity relationship fit to the W11 sample from \cite{Reiners2014} is shown as the dashed grey line, our orthogonal distance fit to the same W11 data is the solid black line. Our target stars cover a wide band of activity levels, and about half of our sample stars cluster along or even below the extreme low-activity edge of the W11 sample (stars 1, 2, 4, and 8).}
 \label{fig:sample_wright}
 \end{centering}
\end{figure}

We then placed the stars into the context of the rotation-activity relationship, which is well-studied for cool stars of various masses and ages. We chose the comprehensive sample presented by W11 who give the luminosities of their sample stars in the ROSAT energy band of 0.1-2.4~keV, and we select the calculated X-ray luminosities of our stars in the same band for comparison. 

While W11 presented their sample in terms of the Rossby number, i.e.\ the stellar rotation period divided by the convective turnover timescale $\tau_c$, we have chosen an approach based purely on the stellar rotation period and X-ray luminosity as presented by \cite{Reiners2014} (see section~\ref{sec:methodology:sample}).  We note that \cite{Reiners2014} calculated a trend line for the W11 sample through standard linear regression; however, for samples in which both x and y experience significant scatter, an orthogonal distance regression (ODR, \citealt{Boggs1990}) is more appropriate (see also \citealt{Hogg2010}). We re-calculated the trend line for that sample through ODR with the scipy.odr package\footnote{\url{https://docs.scipy.org/doc/scipy/reference/odr.html}} and display both our trend line and the one from \cite{Reiners2014} in Fig.~\ref{fig:sample_wright}; our re-calculated line is described by the formula $\log L_X [\mathrm{erg\,s^{-1}}] = 30.71 - 2.01 \log P [d]$.

For completeness, we also give a representation of our sample in the context of the W11 data in the more traditional $L_X/L_{bol}$ versus $Ro=P/\tau_c$ plot in the Appendix, where we have estimated the convective turnover timescales for our sample stars from $V-K$ stellar colours as prescribed in W11 (see Fig.~\ref{fig:sample_wright_Rx_Ro}). 

Returning to Fig.~\ref{fig:sample_wright}, we find that the sample stars cover a rather wide band in $L_X$ activity levels, with two of the upper limits sitting near the lower edge of that range, meaning that the true activity level distribution is even wider. Several observations can be made from the plot: 

1) The sample stars occupy a space that is only poorly covered by the W11 sample. This is not surprising, since we specifically targeted old sun-like stars with deep X-ray exposures, which would otherwise not have been detected in shorter observations. 

2) The distribution of the sample stars around the trend line is significantly different from the other W11 stars in this rotation period range. For W11 stars with $P_{rot} \geq 10$\,d, we see that our sample extends to lower X-ray luminosities at a given rotation period than what is displayed by the W11 sample. In particular, the X-ray detected stars 51~Peg (star 8) and KIC~6106415 (star 1) as well as the two X-ray upper limits for KIC 6116048 (star 2) and KIC 10963065 (star 4) deviate farther from the scatter around the trend line of the W11 sample than expected. 

3) The position of the sample stars in the rotation-activity diagram correlates with their deviation from the rotational behaviour of their M67 counterparts. We have two stars that rotate more slowly than their M67 counterparts, i.e.\ they behave normally for their age in rotational behaviour; they are denoted by red symbols. Those stars sit close to the activity trend line of the W11 stars, however one of the data points (star 5) is an upper limit. The rest of the stars are ones that rotate faster than their M67 counterparts, denoted by blue symbols. Those stars all sit below the trend line of the W11 sample.

\begin{figure}
\begin{centering}
 \includegraphics[width=0.49\textwidth]{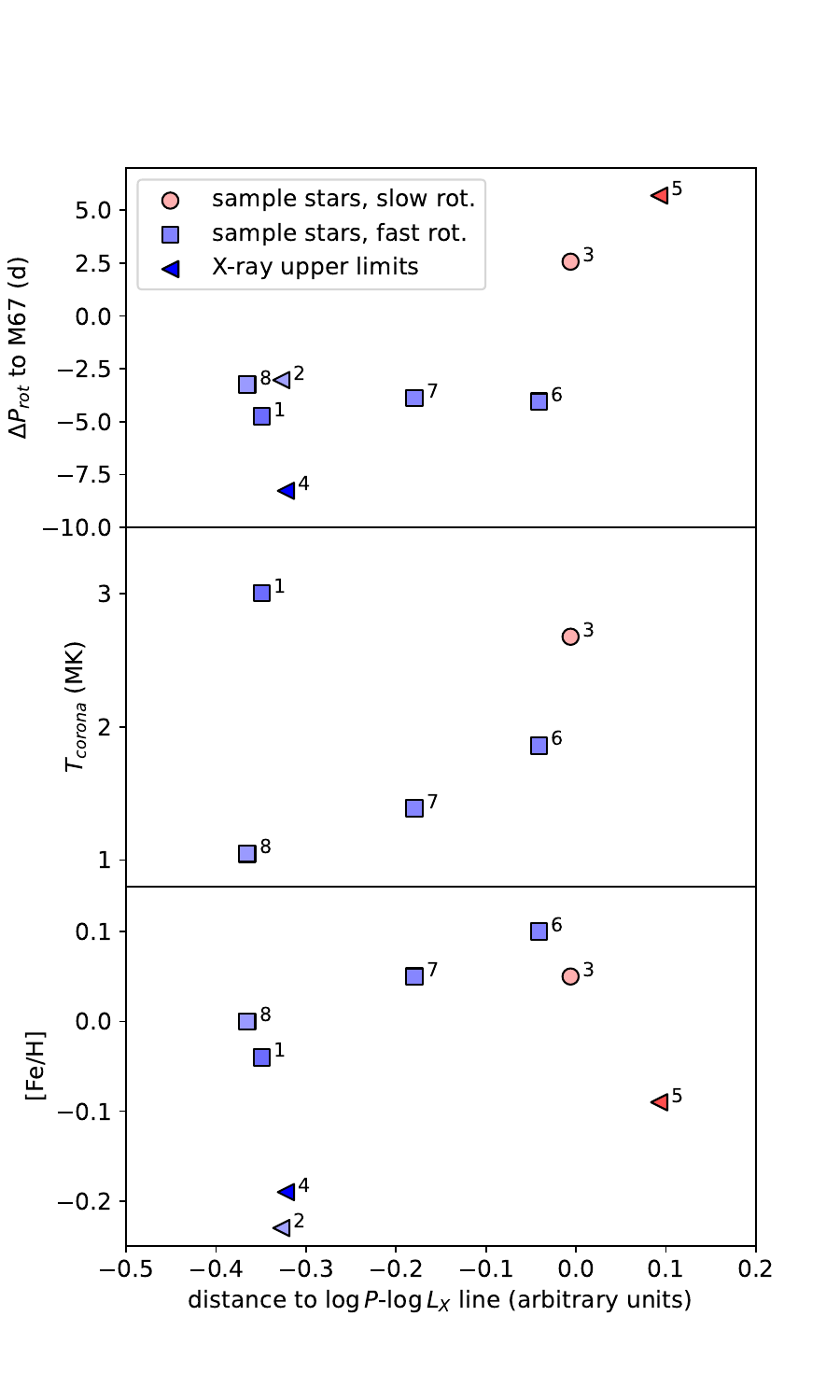}
  \caption{Several properties of the sample stars shown as a function of each star's distance to the canonical rotation-activity relationship from Fig.~\ref{fig:sample_wright}. Due to the limited sample size, the potential trends discussed in the text are to be taken as suggestive, but not as statistically significant. Top panel: The stars that rotate faster than their counterparts in M67 despite being older are also the ones that are located farthest below the rotation-activity relationship. Middle panel: For the stars where we could constrain coronal temperatures, there is potentially a trend of decreasing coronal temperature with stronger disagreement to the rotation-activity relationship, if star 1 is an outlier. Bottom panel: Three stars show significantly sub-solar metallicity ($\lesssim 0.1$\,dex, triangle symbols); those are the stars we did not detect in X-rays and were only able to derive upper limits on their X-ray luminosity for.}
 \label{fig:orthodist}
 \end{centering}
\end{figure}

We investigate this deviation from the W11 trend further by placing the orthogonal distance to the trend line, i.e.\ to the solid line in Fig.~\ref{fig:sample_wright}, into context with different stellar activity and rotation observables. In Fig.~\ref{fig:orthodist}, we display the rotation difference to M67, the inferred coronal temperature for that stars that were detected in X-rays, and the stellar metallicity from \cite{Hall2021} as a function of orthogonal distance to the rotation-activity-relationship. 

We find that the four stars that are deviating the most from the W11 trend line (stars 1, 2, 4, 8) display some unusual properties compared to the rest of our sample. Stars 2 and 4 have significantly sub-solar metallicities, and only have X-ray upper limits. Star 8 has the lowest coronal temperature estimate of the sample (note that we can only determine coronal temperatures for X-ray detected stars, i.e.\ we do not have information about the coronal temperatures of the stars 2 and 4). Star 1 is the only star with large distance to the W11 trend line that does not have suspiciously low metallicity or a very low coronal temperature.

\section{Discussion}\label{sec:disc}

Our stellar sample is quite small with only eight stars, because X-ray measurements are very expensive in terms of exposure time for old main-sequence stars, even if they are located in the immediate solar neighborhood ($d<100$ pc). While statistical inferences are challenging to obtain from such a small sample, we still can use the dearly obtained X-ray properties of the sample to gain some insights into the stellar magnetic activity on a per-star basis.

We note that we do not consider our sample stars to be intrinsically biased for outlier magnetic activity behaviour due to the sample construction. While the amplitude and therefore detectability of asteroseismic modes can be suppressed by high magnetic activity of a star, this effect appears to inhibit asteroseismic mode detections in stellar samples at activity levels (expressed in stellar photometric variability) that are higher than the Sun's by factors of a few up to one order of magnitude \citep{2011ApJ...732L...5C}. Such high activity levels are not expected for solar-age, solar-mass stars. And indeed, our sample does not scatter around solar activity, but rather displays below-solar activity on average as shown in Fig.~\ref{fig:sample_wright}, therefore inhabiting an area of the parameter space that does not appear to be intrinsically bound by the detectability of asteroseismic modes.

In this section we will first discuss how reliable the published determinations of the age and rotation of the sample stars are; then we will move on to discussing the results in the context of open clusters and stellar rotational evolution.

\subsection{Are the age and rotation determinations of the target stars reliable?}\label{age_rot_reliability}

\begin{figure*}
 \includegraphics[width=0.49\textwidth]{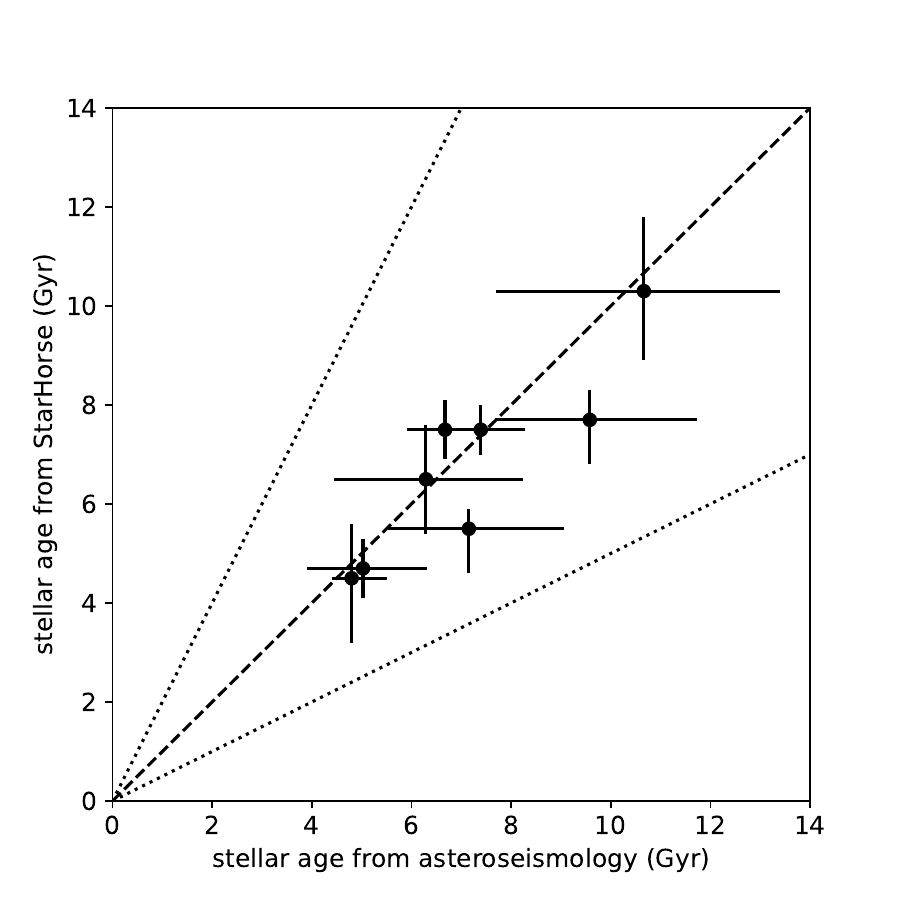}
\includegraphics[width=0.49\textwidth]{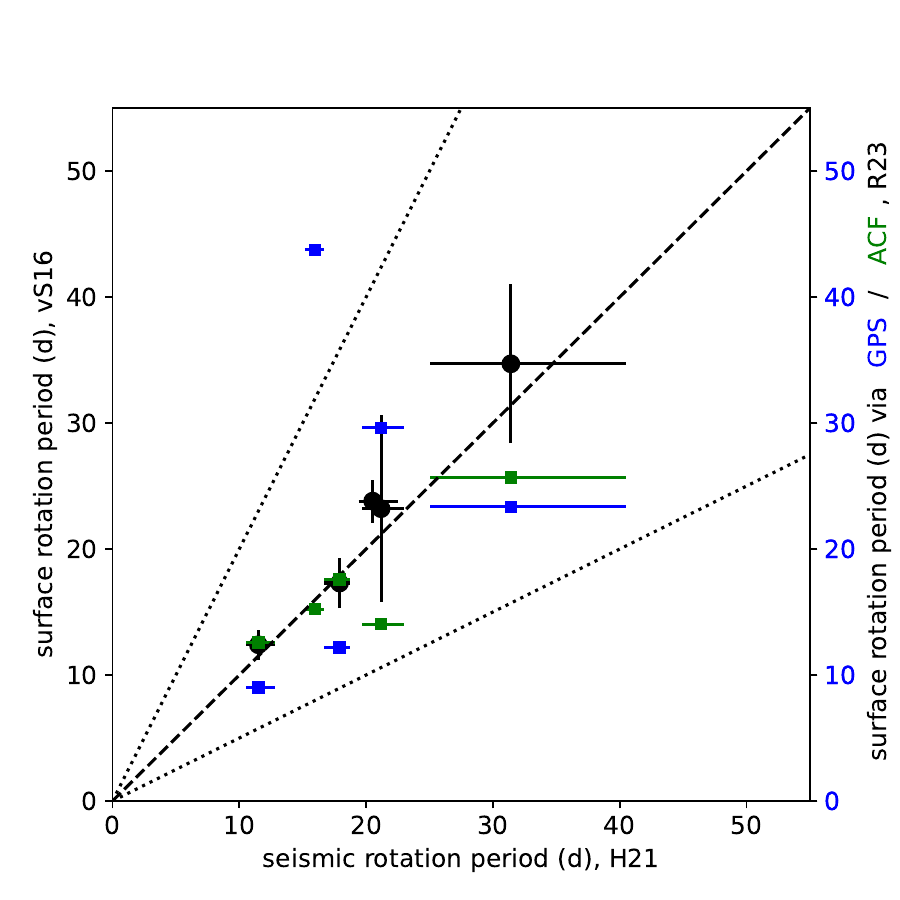}
  \caption{\textit{Left panel:} The stellar ages of the target stars as estimated from asteroseismology as given by \cite{Hall2021} and from the isochrone-based {\tt StarHorse} algorithm do not differ significantly; see text for details on the given StarHorse uncertainties. The dashed line represents equal age estimates, the dotted ones a factor of two between the age estimates. \textit{Right panel:} The seismic rotation periods of the sample stars from \cite{Hall2021} (H21) agree well with the rotation periods from \cite{vanSaders2016} (vS16), which were mostly derived from lightcurve modulation (black filled circles). The rotation periods from \cite{Reinhold2023} (R23) also use lightcurves, but employ the Gradient of Power Spectra method (blue squares), which may be more reliable for stars with solar-like or lower than solar activity. The R23 periods from traditional ACF are given in green (see text for details). The GPS periods deviate from the seismic periods more strongly overall, and quite significantly for one of the sample stars.}
 \label{fig:age_rot_comparison}
\end{figure*}

Age determinations from asteroseismology fundamentally rely on stellar evolutionary models, since the age estimate is based on the sound speed and therefore density in the stellar interior, which can be inferred from the observed asteroseismic oscillation spectrum \citep[e.g.][]{Chaplin2013}. Asteroseismic age estimates may therefore differ if different underlying models are used (see for example \citealt{Lebreton2014}). Furthermore, it has been demonstrated that stellar activity due to spots can suppress the oscillation signal, so that the detectability of asteroseismic modes can be entangled with stellar activity of the star \citep{Chaplin2011, Bonanno2014}; however, we are dealing with magnetically quiet stars here.

No asteroseismic age studies have yet been performed for main-sequence stars in open star clusters older than the Sun, mainly due to brightness limitations, but also due to the scarcity of such objects in the extended solar neighborhood (e.g. \citealt{Wielen1971, Anders2021}). Therefore, a methodological link in the ``age ladder'' (in analogy to the distance ladder) has not yet been established between the ages derived from open stellar clusters and from asteroseismology for the relevant stellar population we study here.

In the absence of such a link, we can refer to other age estimates for the sample stars and check for overall consistency in age estimates between different methods. 
Here, we employed the Bayesian stellar parameter estimator code {\tt StarHorse} \citep{Queiroz2018, Queiroz2020, Queiroz2023} to derive a different estimate of the stellar ages. The code uses astrometric, spectroscopic, and photometric information to find the joint posterior in distance, extinction, and fundamental stellar parameters (age, mass, and metallicity) based on a fine grid of PARSEC 1.2S stellar models. For this work, we have used the {\it Gaia} DR3 parallaxes and photometry, broad-band photometry from Tycho-2 \citep{Hog2000} and 2MASS \citep{2MASS}, as well as the high-resolution spectroscopically determined $T_{\rm eff}$ and [Fe/H] estimates and asteroseismically determined $\log g$ estimates with the appropriate uncertainties. Since in our case we have additional constraints on the mass from asteroseismology, we also included the mass estimate for each star. {\tt StarHorse} converged for all eight sample stars. Appendix figure \ref{fig:starhorse_ages} shows our posterior age probability density functions (PDFs) for the sample stars. As expected for stars still on or close to the main sequence, the age PDFs are typically broad, and in some cases multi-peaked. In most cases, however, thanks to inclusion of of the seismic mass as a constraint, the age PDFs are reasonably close to Gaussian and well-constrained with resulting age uncertainties of $\lesssim2$ Gyr. It is also reassuring that 16 Cyg A and B are found to be co-eval ($\tau_{16 Cyg A} = 7.5^{+0.7}_{-0.5}$ Gyr, $\tau_{16 Cyg B} = 7.5^{+0.4}_{-0.4}$ Gyr) and reasonably close to the precision estimate of \citet{Bazot2020} who reported a seismic age of $\tau_{16 Cyg} = (6.88 \pm 0.12)$ Gyr for that system. For the following comparison, we use the 50th percentile of each star's age PDF as the age value, and the 16th and 84th percentile as the lower and upper end of the error bar, respectively. However, we point out that these nominal uncertainties do not include the systematic stellar modelling uncertainties; \cite{Tayar2022} demonstrate that, when taking into account discrepancies between evolutionary models, stellar age uncertainties typically are of the order of $20\%$, i.e.\ significantly larger than the nominal Starhorse age uncertainties. 

We show the comparison of the asteroseismic and the {\tt StarHorse}-derived age estimates in Fig.~\ref{fig:age_rot_comparison}, left panel. The ages agree well overall; while there is some scatter, it is at the level expected from the reported uncertainties. Most importantly, both age estimates for the sample stars place them consistently at ages older than M67, independently of the method used, which is a relevant outcome for our further discussion.

Now we turn to the rotation periods. The rotation periods from \cite{Hall2021} are derived from the asteroseismic oscillation spectrum, i.e.\ they are not necessarily the same as surface rotation periods, which may introduce a systematic difference when comparing to other stellar samples that use surface rotation periods. On the other hand, the advantage of this method is that it should be able to detect rotation also in the absence of any surface spots, which is why we chose the \cite{Hall2021} sample as our starting point in the first place.

We compare the rotation periods from H21 to the ones reported by \cite{vanSaders2016}, vS16 from here on, and \cite{Reinhold2023}, R23 from here on. Rotation periods from vS16 are present for five out of the eight sample stars, but only three of those stem from rotational surface modulation in light curves; for 16 Cyg A and B, also vS16 report a seismic rotation period determination. Nevertheless, as shown in the black large dots in Fig.~\ref{fig:age_rot_comparison}, right panel, the periods agree very well overall, without any apparent significant difference. 

The rotation periods from R23 are derived also from surface modulation in light curves, but they use a different framework inspired by how solar rotational variability looks like in the Sun's Total Solar Irradiance light curves. Essentially, photospheres of low-activity stars like the Sun are known to be dominated by faculae instead of spots \citep{Lockwood2007, Montet2017, Reinhold2019}, which have a different spatial extent on the stellar surface than spots and were shown to be more reliably detected by analyzing the gradient of the power spectrum (GPS) of the stellar light curve and search for an inflection point \citep{Shapiro2020, Amazo-Gomez2020a, Amazo-Gomez2020b, Reinhold2023}. R23 use both the GPS method and the more traditional autocorrelation function (ACF) method, such as used by \cite{McQuillan2014}, to probe their stellar light curves for rotation. For our sample stars, R23 reports the GPS-derived period as the more reliable rotation period estimate in all cases. Those rotation periods differ quite significantly from the H21 and vS16 periods in some cases, as shown in Fig.~\ref{fig:age_rot_comparison} in blue colour. The R23 periods determined from ACF (green color) agree reasonably well with the asteroseismic ones and the surface rotation periods from vS16. For completeness, we display the stellar activity of the sample stars in the context of their R23 GPS periods in the Appendix, Fig.~\ref{fig:rot_act_R23}.

We therefore proceed using the asteroseismic rotation periods from H21, but point out some subtleties in the different rotation period determinations in section~\ref{disc:m67}, in particular in the context of rotation periods in stellar clusters. Those subtleties could be investigated further in samples larger than the one we have available here.

Overall, we argue that neither the age estimates nor the rotation period estimates are wrong in a systematic way that would explain why the sample stars sit at unexpected places in age-colour-rotation space and in the rotation-activity plane.

\subsection{No dynamo breakdown, but a mix of truly weak activity and low metallicity}

The stars in our sample can be loosely put into two groups, with four stars sitting close to the regular rotation-activity relationship displayed in Fig.~\ref{fig:sample_wright} (stars 3, 5, 6, 7) and four stars sitting at the outer low-activity edge of the spread spanned by the W11 sample (stars 1, 2, 4, 8). Keeping in mind that all our sample stars have age estimates older than M67, we come back to 
Fig.~\ref{fig:orthodist}, where we show different stellar properties as a function of each star's orthogonal distance to the empirical W11 rotation-activity relationship. 

The former group of sample stars does not display unusually low activity for the stellar rotational state; there is, however, one rather unrestrictive X-ray upper limits among them. Two of those stars are actually spun down compared to stars of similar colour in M67. The other two stars have a slight over-rotation for their age and normal activity for their rotation (16~Cyg~A and B, i.e.\ stars 6 and 7; again note that the unresolved M~dwarf companion to 16~Cyg~A is extremely X-ray faint and therefore not an issue in this, see Appendix~\ref{appendix:16cygc}). 

The latter group of very low-activity stars is characterized by two X-ray upper limits and two X-ray detections. Both upper limits here stem from stars with a low metallicity of [Fe/H]$\approx-0.2$, see stars 2 and 4 in the lower panel of Fig.~\ref{fig:orthodist}. For stars with low metallicity, it is not surprising that they behave differently in their rotational evolution compared to solar-metallicity stars \citep{See2024}. Recent works to determine convective turnover timescales empirically have found a distinctive influence of stellar metallicity  in which the effective turnover timescale is shorter for stars with low metallicity, leading to larger Rossby number at the same rotational period; see for example \citealt{Bonanno2025}, who explicitly demonstrate this not just when comparing stars of similar mass, but also stars of similar \textit{Gaia} color. This agrees with theoretical works in which convective turnover timescales are inferred from stellar models, which also predict shorter timescales at lower metallicity (see for example \citealt{vanSaders2013}). In this sense, a G-type star with sub-solar metallicity would behave more like an F star than its same-age, same-mass cousin with solar metallicity. We try to quantify this effect roughly by comparing the empirically calibrated convective turnover timescales from \cite{Bonanno2025} for a 5\,Gyr solar-mass star with either solar or sub-solar metallicity of $\mathrm{[Fe/H]}=-0.2$\,dex. Equation~16 of \cite{Bonanno2025} estimates convective turnover timescales $\tau_{conv}$ from stellar Gaia colour, effective temperature, surface gravity, and metallicity. For the solar-metallicity version, we use solar values for those quantities, and estimate the respective quantities for the low-metallicity star from stellar evolution models\footnote{This exercise yields differences of the lower-metallicity star being roughly 230\,K hotter, 0.05\,mag bluer in Gaia Bp-Rp colour, and 0.04 dex lower in $\log g$, using PARSEC evolutionary tracks \citep{2012MNRAS.427..127B}.}. Comparing the resulting convective turnover timescales, the low-metallicity star's $\tau_{conv}$ is about 35\% shorter than the solar-metallicity one. This roughly matches the $\tau_{conv}$ of a solar-metallicity G0-F9 star. However, since we already present our sample in terms of Gaia colour and not mass (for example in Fig.~\ref{fig:hyperplane} and \ref{fig:m67rotation}), the metallicity-induced effective temperature and colour difference should not skew this comparison. The remaining effect of the pure metallicity and the metallicity-induced surface gravity difference on $\tau_{conv}$ is rather small in the \cite{Bonanno2025} formulation; it accounts only for a ca.\ 10\% decrease of the convective turnover timescale compared to the solar-metallicity star. In an instantaneous view, this may not fully explain why sample star 4, for example, is rotating almost twice as fast as a similar-colour M67 star; but in an evolutionary view, this may have added up to a reduced magnetic braking overall.
Further taking into account that the rotation period sequence of stars in middle-aged open clusters like Ruprecht-147 and M67 drops quite sharply towards shorter rotation periods when going to the F stars regime at bluer $(BP-RP)_0$ colour (see Fig.~\ref{fig:m67rotation}) and is not very well sampled, the rotational state of our two low-metallicity stars may be less out of line with what is seen in open clusters than it appears.

For the two other stars in the very low activity group, namely 51~Peg (star 8) and KIC~6106415 (star 1), the situation is different. KIC~6106415 is similar to solar metallicity and 51~Peg is of mildly supersolar metallicity, so their moderately fast rotation at very low X-ray activity is not related to a low-metallicity effect on turnover timescales. 51~Peg has been studied repeatedly in the past, with different X-ray activity estimates either derived from narrow-band X-ray photometry with XMM-Newton \citep{Poppenhaeger2009} or estimates from Chandra's HRC-I instrument, which does not have an intrinsic energy resolution \citep{Metcalfe2024}. Our re-analysis here, using also the very soft X-ray photons in the 150-200~eV range, leads to a slightly lower estimate of the overall X-ray luminosity of this star than previous works. It is remarkable that both 51~Peg and KIC~6106415 are characterized by a very low X-ray flux through the stellar surface of $\lesssim 10\,000$\,erg\,s$^{-1}$\,cm${-2}$ in the 0.1-2.4~keV energy band, see also Table~\ref{tab:results}. An X-ray flux threshold of $10\,000$\,erg\,s$^{-1}$\,cm${-2}$ has been identified by \cite{Schmitt1997} as an empirical lower boundary to the X-ray activity of stars in the solar neighborhood; this value was reported by \cite{Schmitt1997} to coincide with the the typical X-ray surface flux of a solar coronal hole. We point out that at least in the case of 51~Peg there is a consistently low X-ray luminosity observed since the 1990s with ROSAT, meaning that we are likely not witnessing a temporary low activity state due to a stellar activity cycle here.

Interestingly, the derived coronal temperatures for those two stars are quite different -- 51~Peg displays a very low coronal temperature of about $1$\,MK, which aligns with the expectation that stars with low X-ray surface flux have low coronal temperatures. This is well-established among cool stars \citep{Telleschi2005, Schmitt1997, Preibisch1997, Johnstone2015}. In contrast, KIC~6106415 has a temperature of about $3$\,MK, which is typically seen for stars with moderate activity level. This star is an outlier in our admittedly small sample of coronal temperature determinations, as shown in both Fig.~\ref{fig:hardnessratio_activity} and Fig.~\ref{fig:orthodist}, middle panel. A possible explanation might be that a small flare occurred during the roughly half-day long exposure, with flares displaying higher coronal temperatures than the quiescent or quasi-quiescent coronal emission; unfortunately, the low number of excess X-ray photons detected does not allow for a variability analysis.

It is worth mentioning that very low coronal temperatures could be interesting in terms of inhibited angular momentum removal from the star. For a very low temperature corona it becomes difficult to launch a wind with sufficient velocity to escape the gravitational well of the star. Different scenarios have been explored in the literature; for example, the stellar wind mass-loss rate might be reduced, potentially causing a drop in angular momentum loss at low coronal temperatures \citep{OFionnagain2018}. However, \cite{Pantolmos2017} explored the effect of coronal temperature and wind speed in combination with the Alfv\'en radius, finding the most relevant parameter to be the average wind speed at the Alfv\'en surface in order to compare angular momentum losses for different coronal scenarios. \cite{Levesque2025} explored winds for stellar coronae at low temperatures close to $1$\,MK, finding that magnetocentrifugal effects may reduce the Alfv\'en radius and lead to strongly reduced angular momentum loss rates even if mass loss rates were not much different from the ones at higher coronal temperatures. 

In the context of an emerging picture of stellar spin-down beyond the solar age in the recent literature, we consider this work to provide missing information about small-scale stellar coronal structure and indications about stellar winds launched from there. For a star to spin down via magnetic braking, a corona heated by small-scale magnetic structures needs to be in place, from where a stellar wind can be launched; the wind then carries away angular momentum along the larger-scale stellar magnetic field. 
\cite{Metcalfe2025ApJ...991L..17M} argue for a collapse of specifically the large-scale magnetic field of older sun-like stars as an explanation for their observed over-rotation, together with a potentially reduced mass loss rate.  Our work demonstrates that small-scale magnetic activity is generally present for our stars, since we detect coronal X-ray emission for most of them. A complete dynamo collapse at all spatial scales is therefore not indicated. Based on the low coronal temperatures observed for part of our sample, one may speculate that wind launching from such coronae is weakened and leads to low angular momentum loss. This finding is neither in support of nor contradiction to potentially  collapsed large-scale magnetic field structures for such stars: 
A reduced angular momentum shedding for these stars could arise from insufficient wind launching from cool coronae, from a weakened large-scale magnetic field, or from both.

\subsection{Where are the fast-rotating suns in M67?}\label{disc:m67}

The majority of the targets in our sample display moderately fast rotation at an asteroseismic age estimate older than the solar age. Even if we ignore the two stars with low metallicity as being non-representative of G star behaviour as discussed above, we still have four stars that rotate faster than stars of the same colour in M67, despite having a significantly older age estimate (these are the sample stars with Gaia $(BP-RP)_0 \leq 0.9$\,mag depicted in blue in Fig.~\ref{fig:m67rotation}). This remains true even if one considers a certain spread in the majority of the published cluster age estimates for M67 (see section~\ref{sect:age-rot}).

The weakened magnetic braking scenario predicts that sun-like stars enter a regime where magnetic braking does not work efficiently any more, leaving them with a more or less constant rotation period after entering this regime at a certain age. In principle, the very weak activity we observe for the over-rotating part of our sample might match an insufficient magnetic braking scenario. If we stipulate that those stars are indeed in the weakened magnetic braking regime, they would have needed to enter this regime at an age somewhere between the age of Ruprecht~147 ($\sim2.7$~Gyr) and M67 ($\sim4$~Gyr), based on their current rotation periods. This poses the question: why have the stars in M67 spun down more than our blue-coloured sample stars, despite M67 being younger? Either this is a real physical phenomenon that causes field stars to evolve differently in their rotational behaviour compared to cluster stars, or there are subtle effects in sample selection, rotation diagnostics, or age calibration that cause an apparent difference. We discuss both options in turn.

If the observed rotation difference between our sample stars and the M67 stars is real, then the question is which physical mechanism could cause that difference. The observed sample behaviour is not straightforward to interpret under the assumption of a scenario in which weakened magnetic braking is the only effect in play. Recent work suggests that the onset of weakened magnetic braking for more massive stars should happen at shorter rotation periods and younger ages than for less massive stars \citep{Saunders2024}, which would in principle be in line with the colour-dependent behaviour we see in our sample. However, if that onset for the blue stars in the sample truly were to happen somewhere around an age of 3~Gyr, then the M67 stars should be subject to the same effect, and we should not see them rotating more slowly than our sample stars.
An additional scenario could be the redistribution of angular momentum inside the star, leading to an effective spin-up of the outer stellar layers. Such an effect would need to become sufficiently strong at ages older than M67 and younger than our sample stars, i.e.\ between 4 and 5 Gyr, to explain the rotational discrepancy. However, while redistribution of angular momentum inside cool stars has been suggested in some works \citep{Curtis2020, Spada2020, Santos2025}, the ages at which this is expected to happen for solar-mass stars \citep{Spada2026} are much younger than the observed age range we study here, making such a redistribution unlikely to explain the observed rotational behaviour.

The other option is that sun-like field stars beyond the M67 age do not rotate faster than the M67 stars, and the observed mismatch is due to observational biases. We argue that \textit{basic} observational biases are unlikely to explain this. Regarding the rotation measurements, the rotation periods of the cluster stars in Fig.~\ref{fig:m67rotation} were determined from the rotational modulation of light curves, i.e.\ represent surface rotation periods. In contrast, the asteroseismic period determination from H21 probe a deeper layer of the star, which could lead to a different rotation period due to differential rotation in the radial direction. However, as we have shown in Fig.~\ref{fig:age_rot_comparison}, for the sample stars that have both surface rotation periods from \cite{vanSaders2016} and seismic ones from \cite{Hall2021}, there is no systematic offset between the two, which is consistent with other studies performing similar comparisons (see for example \citealt{Benomar2015}). The over-rotation of the blue stars compared to M67 is therefore likely real. We have also discussed the reliability of the inferred stellar ages in section~\ref{age_rot_reliability}, which are consistent across two different methods, so the sample stars are likely truly older than M67.

We therefore suspect that the discrepancy most likely comes from a combination of actual subtle changes that happen to sun-like stars beyond the M67 age together with the specifics of the rotation period determinations. One aspect is that could be age-related changes in differential rotation not just radially, but also latitudinally on the stellar surface, that come into play. As mentioned above, asteroseismology represents an averaged view of the rotation in the stellar interior, mainly in the near-surface layers. Light curve modulation by star spots represents the rotation period at the stellar surface, in particular at the stellar latitudes where spots or active regions typically appear for a given star. Depending on the differential surface rotation and the dominant spot latitudes of sun-like stars at different ages, detected surface rotation periods could be slightly different from the asteroseismically determined ones already through this effect.

Another rotational aspect is that for rotation periods from stellar surface brightness modulation different techniques can yield different results. Low-activity sun-like stars are actually expected to be dominated by bright faculae and not dark spots in their rotational light curve modulations, and recent years have highlighted that a network of bright faculae is hard to detect with classical Lomb-Scargle or autocorrelation analyses (see also section~\ref{age_rot_reliability}). Specifically, for two of our sample stars R23 find significantly longer periods through the GPS technique in comparison to the ACF technique, namely for 16~Cyg~B and KIC~6106415. Their work determines that GPS should be more reliable for those stars than ACF; however, the ACF periods agree much more closely with the (shorter) asteroseismic rotation periods. Concerning surface rotation periods in stellar clusters, \cite{Gruner2023} determine their rotation periods in M67 not only via ACF and Lomb-Scargle periodograms, but show that the technique of phase dispersion minimization \citep{Stellingwerf1978}, which picks up also non-sinusoidal, multi-peaked signals, yields the most consistently reliable results for their sample. Both \cite{Gruner2023} and \cite{Curtis2020} emphasize the importance of weeding out half-periods in vetting the picked up periods, particularly in case of multi-peaked signals. However, for our sample stars it is not obvious how half-periods could be at fault here; the asteroseismic periods should not be that drastically different from surface rotation, at least based on expectations from the solar differential rotation.

To understand how sun-like stars truly evolve in their rotation at sun-like ages and beyond, fully understanding this apparent discrepancy between field and cluster stars is indispensable. To shed light on this, future rotational studies of even older open stellar clusters (such as NGC 188 or NGC 6791) would be highly desirable, but due to the brightness limitations of recent, current, and near-future space observatories such as \textit{Kepler} \citep{Borucki2010Sci...327..977B}, \textit{TESS} \citep{Ricker2015JATIS...1a4003R} and \textit{PLATO} \citep{Rauer2025ExA....59...26R}, such observations are challenging with current observational means. Alternatively, a careful connection between the asteroseismic age estimates to the ages derived for stars in open stellar clusters would help significantly in improving confidence in asteroseismic ages for cool main-sequence stars. This could be performed with asteroseismic studies of such stars in nearby open clusters with well-determined ages. Furthermore, in-depth comparisons between different surface rotation period determinations will be helpful. Finally, also spectroscopic rotational broadening and chromospheric activity studies of member stars of intermediate-age open clusters could help reveal in a statistical manner whether there is a sub-population of stars that is rotating faster than expected, but displays little to no surface activity that would allow for a rotation period determination via rotational light curve modulation.

\section{Conclusions}\label{sec:con}

We have presented a study of the magnetic X-ray activity of a small, observationally accessible sample of old sun-like stars with asteroseismic age estimates and rotation periods. The majority of the sample stars are known to rotate faster than the Skumanich law suggests. 

We find that the stars do not have a fully collapsed magnetic dynamo; instead, we detect the majority of the stars in the X-ray band in our deep pointings. We do find that the stars with strong over-rotation are also the stars that display very weak magnetic activity. In particular, we find extremely low X-ray surface fluxes below the empirically established lower limit of $10^4$\,erg\,s$^{-1}$\,cm$^{-2}$ in the solar neighborhood. We also find an extremely low coronal temperature of $\approx 1$ million K for the star that deviates the most from the established rotation-activity relationship of \cite{Wright2011}, namely for 51~Peg, which could explain the very low angular momentum loss rate the star may be experiencing.

We struggle to bring our sample into a clear context with the rotation periods observed for the open cluster M67; our sample stars all have inferred ages older than M67. The rotational state of some of our sample stars could in principle be explained by weakened magnetic braking, if the onset of such a weakening happens at a stellar age around $\approx3$\,Gyr. However, it then remains unclear why the stars in M67, at a cluster age of $\approx 4$\,Gyr, have spun down more than the old stars in our sample. Our analysis excludes issues in age or rotation period determinations as causes for the discrepancy. Low metallicities observed for two stars in our sample alleviate this tension somewhat, but cannot fully resolve it.

\begin{acknowledgements}

\renewcommand{\thefootnote}{\fnsymbol{footnote}}

The authors thank the anonymous referee for their insightful comments on the manuscript.

The authors thank Iris Traulsen for discussions and support on handling X-ray detections and upper limits with \verb|edetect_stack|, and Alfio Bonanno for inspiring discussion during the preparation of the manuscript.

KP and NI acknowledge support from the German Aerospace Center (DLR) under project number 50OR2312. KP also acknowledges support from the European Research Council (ERC) under grant agreement 101170037 (ERC-CoG Evaporator).\footnotemark[1]
JvS was supported by the National Aeronautics and Space Administration (NASA) under award No. 80NSSC23K1576. F.A. acknowledges the grant RYC2021-031683-I funded by MCIN/AEI/10.13039/501100011033 and by the European Union's NextGenerationEU/PRTR. This work was partially supported by the Spanish MICIN/AEI/10.13039/501100011033 and by ‘ERDF A way of making Europe’ by the European Union through grant PID2024-157964OB-C21, and the Institute of Cosmos Sciences at the University of Barcelona (ICCUB, Unidad de Excelencia Mar\'{\i}a de Maeztu) through grant CEX2024-001451-M and the project 2021-SGR-00679 GRC funded by the Agència de Gestió d'Ajuts Universitaris i de Recerca (Generalitat de Catalunya).
SAB acknowledges support from the German Science Foundation (DFG) via grant number 573279422.

The authors thank Iris Traulsen for discussions and support on handling X-ray detections and upper limits with \verb|edetect_stack|.

Based on observations obtained with \textit{XMM-Newton}, an ESA science mission with instruments and contributions directly funded by ESA Member States and NASA.

This research has made use of the Unified Cluster Catalogue (UCC; \citealt{Perren_2023}).

\footnotetext[1]{Views and opinions expressed are however those of the author(s) only and do not necessarily reflect those of the European Union or the European Research Council Executive Agency. Neither the European Union nor the granting authority can be held responsible for them.}

\end{acknowledgements}

\bibliographystyle{aa}
\bibliography{bibliography}

\begin{appendix}

\section{From count rates to fluxes: the 16 Cyg AC B system as a high-S/N template}\label{app:fluxes}

Since most of our target stars have relatively few X-ray photons detected, it is important to develop an understanding of the underlying spectral shape of the stellar X-ray emission to convert detected count rates into physical fluxes. For this, we have a closer look at the moderately X-ray bright 16~Cyg triple system, which is part of our sample. The 16~Cyg system consists of the primary G dwarf 16~Cyg~A, which is spatially blended with the tertiary M dwarf 16~Cyg~C in XMM-Newton observations since they are only separated by ca.\ $5^{\prime\prime}$; at a larger separation of ca.\ $40^{\prime\prime}$ from 16~Cyg~A is the secondary G dwarf 16~Cyg~B located \citep{Raghavan2006}. 

We will first demonstrate that 16~Cyg~C is so faint in X-rays that its contribution to the combined 16~Cyg~AC X-ray spectrum can be expected to be negligible; then we present our spectral analysis of 16~Cyg~A and B and demonstrate how we derive useful count-to-flux conversion factors from the stellar hardness ratios.

\subsection{X-ray faintness of 16~Cyg~C}\label{appendix:16cygc}

\begin{figure}
 \includegraphics[width=\columnwidth]{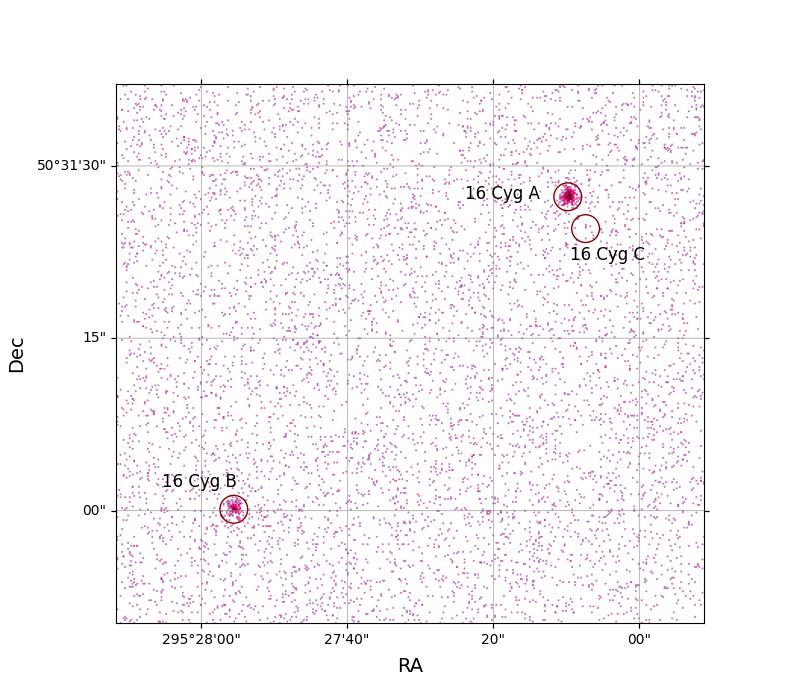}
  \caption{X-ray image of the 16~Cyg~ABC system from two combined Chandra HRC-I observations in the energy band 0.08-10~keV (see text for details). The optical positions of the three stars are marked by circles. The position of 16~Cyg~C is in line with no detectable X-ray flux from the star, meaning that the X-ray emission from the 16~Cyg~AC pair which is spatially unresolved in XMM-Newton observations can be assumed to be practically fully dominated by 16~Cyg~A.}
 \label{fig:16cygChandra}
\end{figure}

Concerning the X-ray flux of 16 Cyg C, we note that spatially resolved X-ray observations of the 16 Cyg triple system exist with \textit{Chandra} (ObsIDs 23188, 21167, 18756, 16647, PI Guinan; an additional ObsID 12740 unfortunately just misses 16 Cyg in its Field of View). 
A detailed analysis of that data is beyond the scope of this paper, but we use the high spatial resolution provided by Chandra to perform a rough estimate of the flux contribution of 16~Cyg~C to the 16~Cyg~A and C pair, which is unresolved in XMM-Newton and resolved with Chandra. Chandra performed two observations with the ACIS-I instrument and two with HRC-I. ACIS-I has practically no energy sensitivity below 0.5~keV, while HRC-I is sensitive to an energy range of 0.08-10~keV, however without any intrinsic energy resolution. Since our XMM-Newton analysis of the unresolved 16~Cyg~A and C pair already shows that their combined emission is dominated by soft X-rays below 0.5~keV, we focus on the HRC-I observations to constrain the flux contribution of 16~Cyg~C. We merged the two HRC-I observations with the \verb|merge_obs| command of Chandra's data analysis software CIAO \citep{ciao}, and display the combined image for the full 0.08-10~keV energy band in Fig.~\ref{fig:16cygChandra}. The expected positions of 16~Cyg~A, B and C, derived from Gaia DR3 data propagated to the epoch of the Chandra observations, are marked as circles. 16~Cyg~A displayed 329 counts in a $1.5^{\prime\prime}$ circular source region over an estimated background of ca.\ 19 counts, while the 16~Cyg~C position is in line with no detectable flux from the tertiary star. We therefore regard 16~Cyg~C to be of negligible contribution to the extracted XMM-Newton data and assign the detected XMM-Newton signal fully to the primary 16~Cyg~A.

\subsection{Spectral analysis of 16 Cyg A and B as a template for deriving coronal temperatures from hardness ratios}\label{appendix:16cygc}

\begin{figure*}
 \includegraphics[width=0.33\textwidth, angle=270]{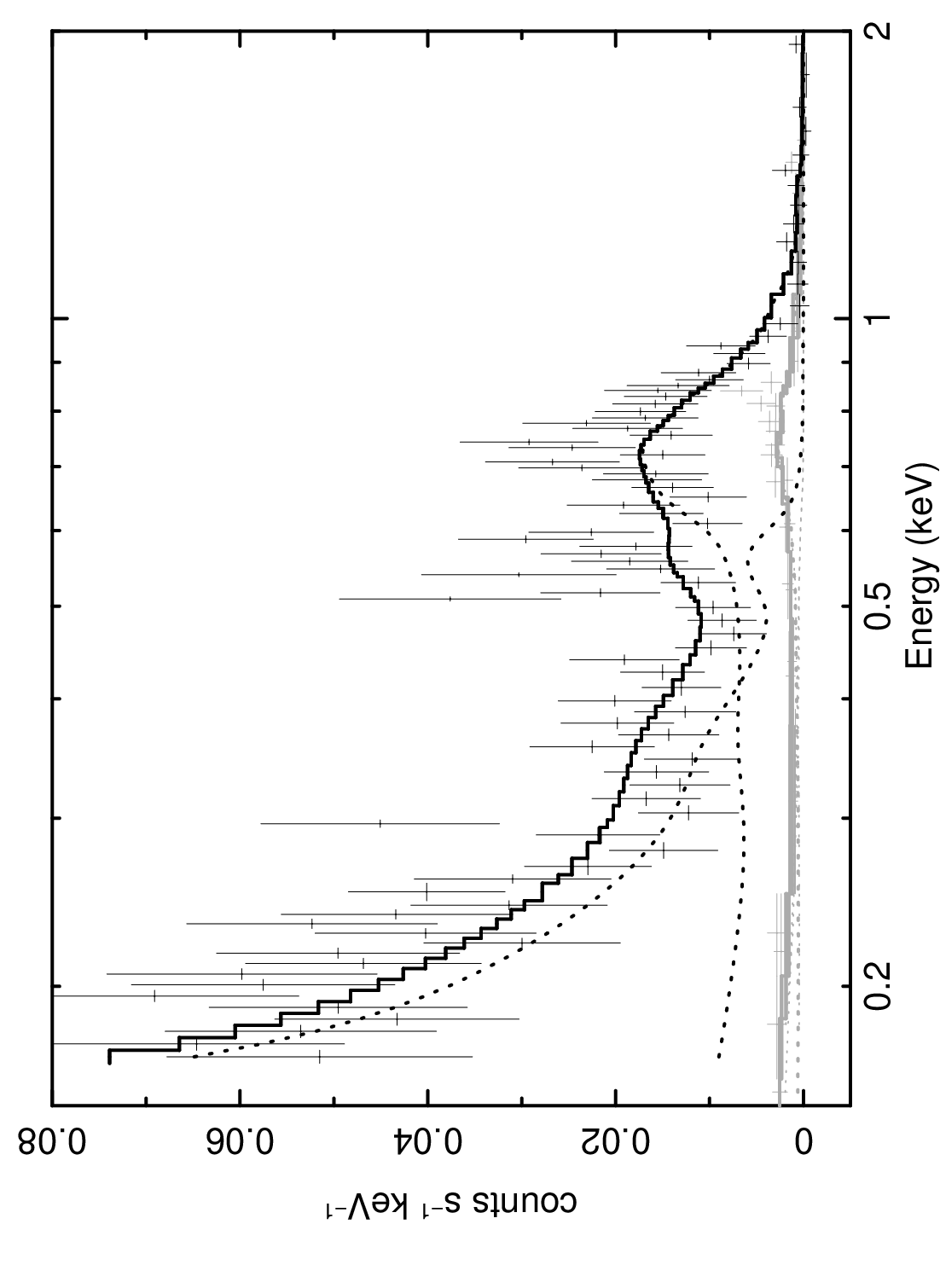}
 \includegraphics[width=0.33\textwidth, angle=270]{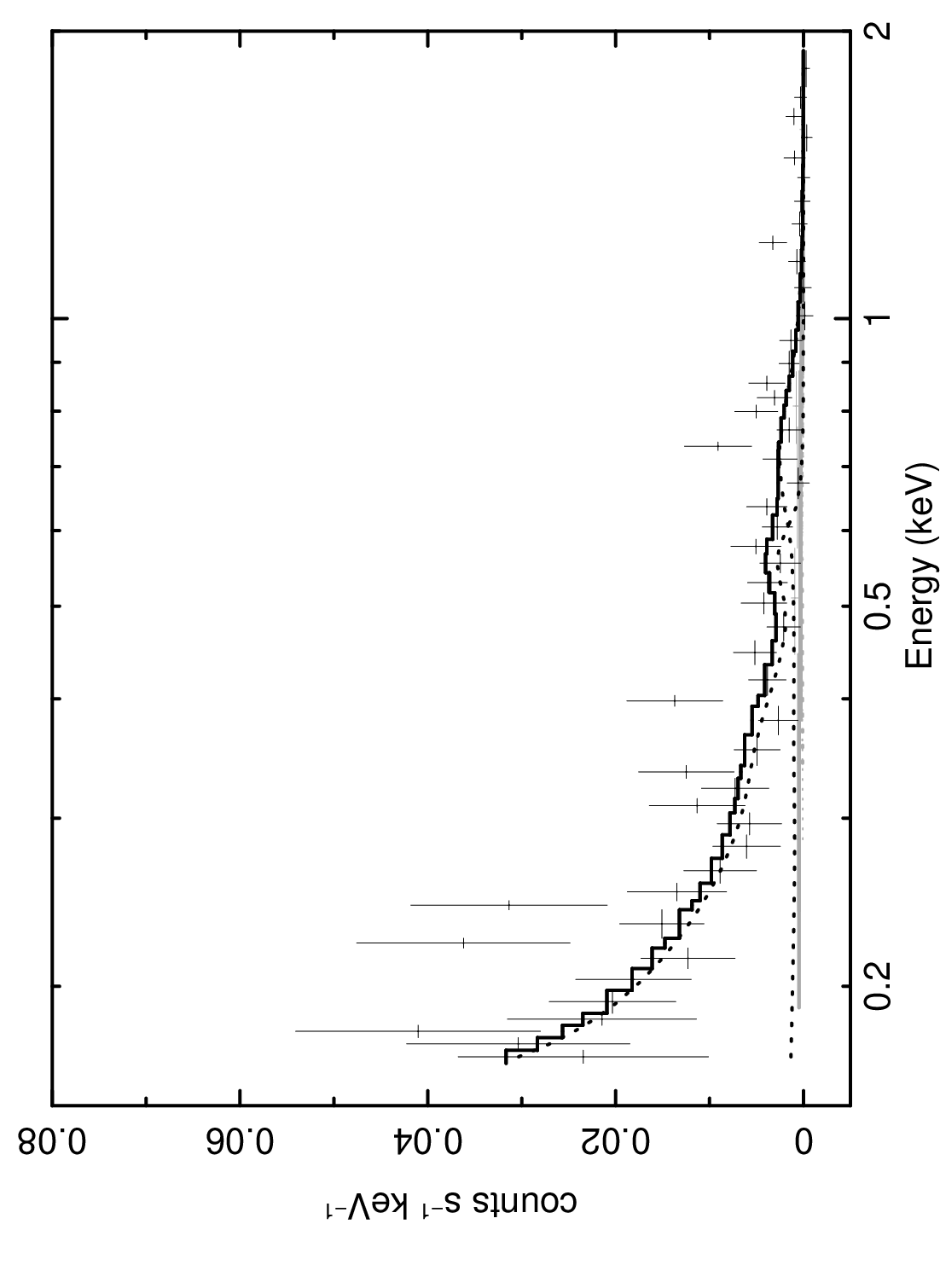}
 \caption{The extracted EPIC CCD spectra of the spatially unresolved 16~Cyg~AC system (left) and the 16~Cyg~B component (right) are very soft. Shown are the PN spectra and the model fit in black and the MOS1 and MOS2 spectra, having much fewer counts, with the same model fit applied in grey. The two temperature components of the model are indicated by dotted lines. The spectra of both AC and B show dominant emission at very soft energies, plus weaker emission in the 0.5-0.9~keV range where the OVII and OVIII lines as well as the iron line complexes appear in stellar coronal spectra. }
 \label{fig:16cygspec}
\end{figure*}

We display the extracted spectra of 16 Cyg A and B individually in Fig.~\ref{fig:16cygspec}. We used the Xspec software \citep{xspec1996}, version 12.14.0, to fit the spectra with optically thin thermal plasma models, as appropriate for stellar coronae. Specifically, we used the \texttt{apec} model and assumed solar abundances from \cite{GrevesseSauval1998}. A single-temperature fit did not yield satisfactory results for either of the two stars. Two-temperature fits yielded excellent fits. It turned out that a two-temperature fit with freely fitted temperatures was actually not better than fixing coronal temperatures on a two-point grid with a mild temperature component at $kT=0.1$~keV and a hot component at $kT=0.5$~keV and letting the mixing ratio in the form of emission measures vary freely. We therefore chose the fixed temperature grid from here on, since it has fewer free parameters and gives us a better shot at applying it also to the other program stars where we have fewer collected X-ray photons.

\begin{figure*}
 \includegraphics[width=0.48\textwidth]{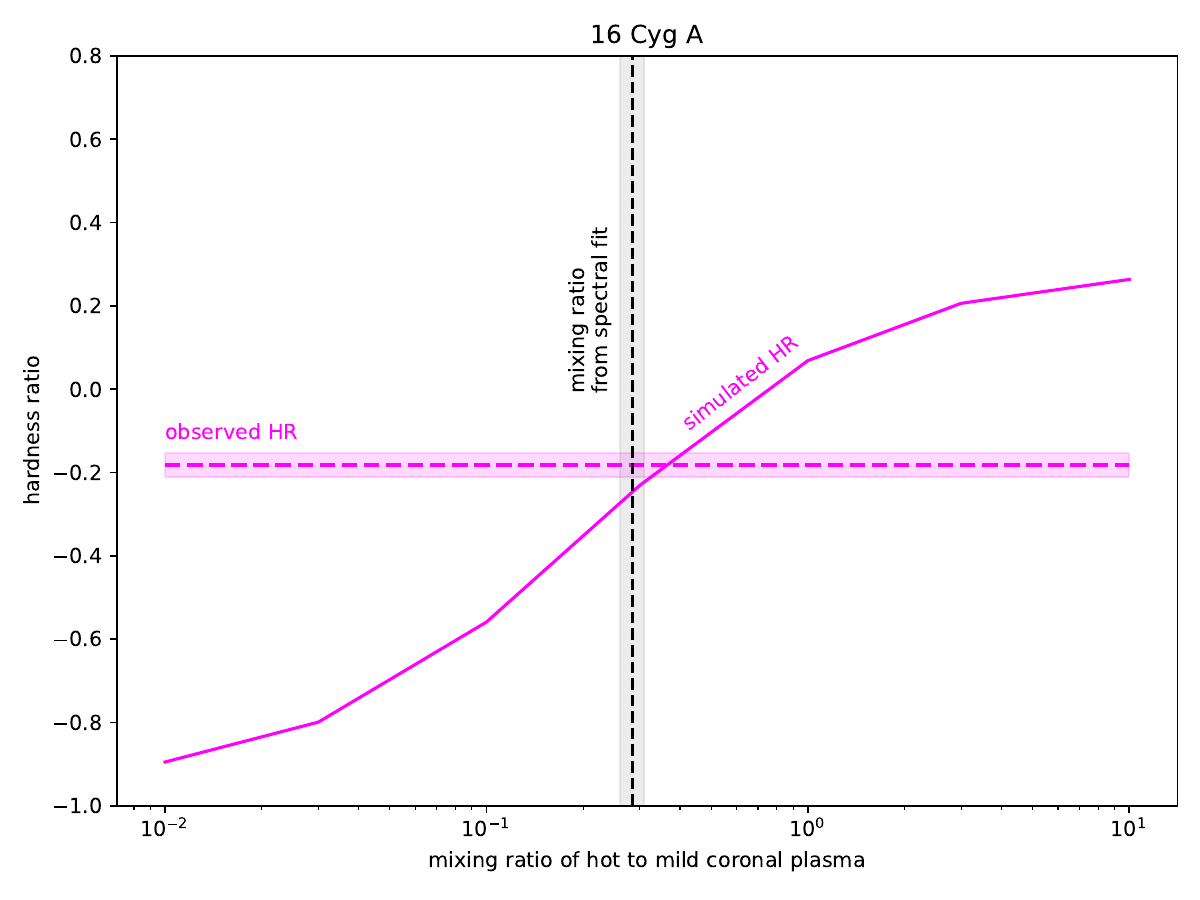}
 \includegraphics[width=0.48\textwidth]{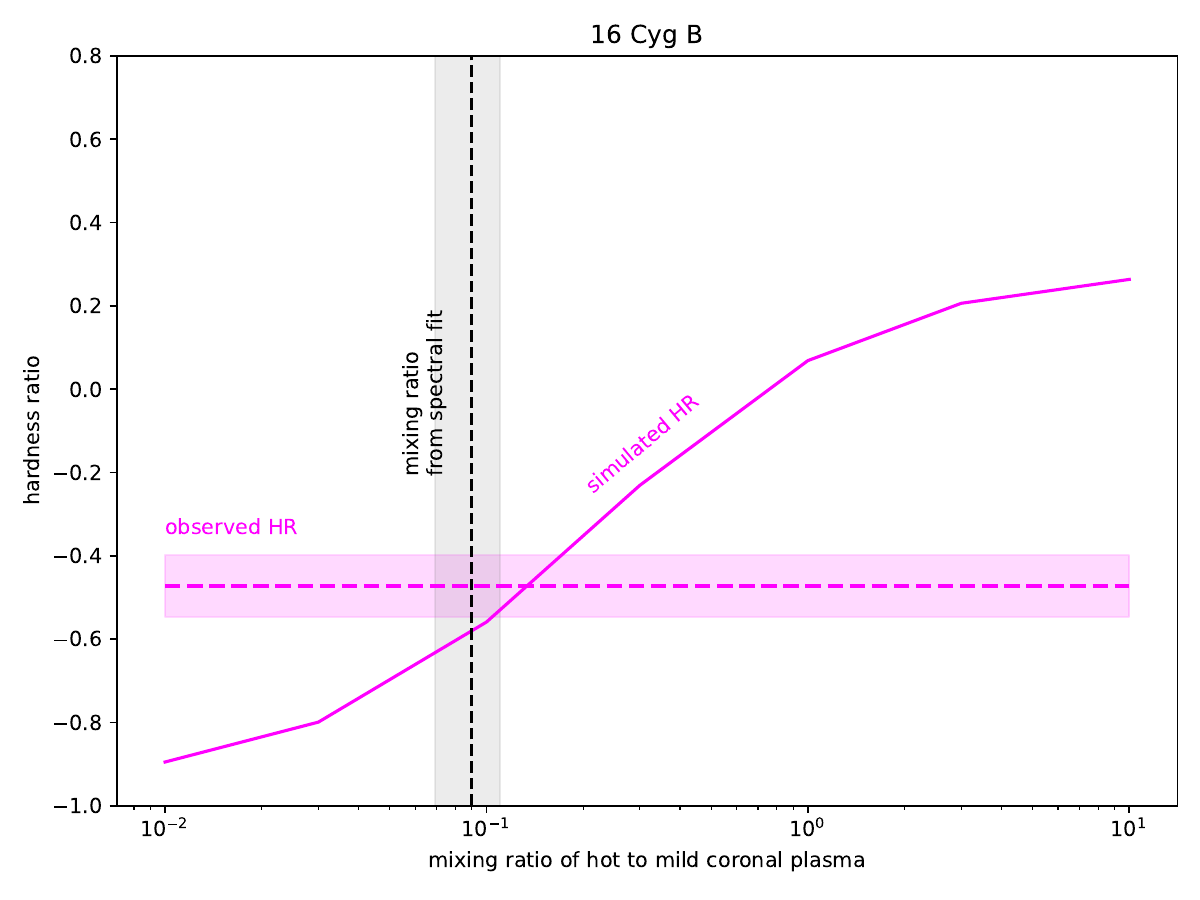}
 \caption{Shown are the simulated hardness ratios for different hot-to-mild coronal plasma mixing ratios, with mild plasma having a temperature of 0.1~keV and hot plasma one of 0.5~keV. The mixing ratio as determined from the spectral fit (see Table~\ref{tab:16cyg}) for 16~Cyg~A and 16~Cyg~B is denoted by the vertical dashed black line; the mixing ratio derived from the measured hardness ratio and its $1\sigma$ uncertainty are displayed by the horizontal magenta band. The spectral and hardness ratio-based estimates of the mixing ratios agree very well. The small offset between them may be due to small uncertainties in the very soft energy calibration of the EPIC instruments.}
 \label{fig:16cygHR}
\end{figure*}

The spectral fits to the 16 Cyg A and B data guide us in our choice for appropriate energy bands for hardness ratios, which can then be used to characterize the coronae of those target stars that do not have sufficient X-ray counts to allow a full spectral analysis. The aim here is to select energy bands in such a way that their hardness ratio allows a good determination of the hot-versus-mild mixing ratio that is present in the corona of the observed target star. For this approach we simulated coronal two-temperature spectra for a range of hot-versus-mild mixing ratios and folded the models with the appropriate spectral response functions for the XMM-Newton EPIC instruments. We then extracted hardness ratios for different energy bands from the simulated spectra and plotted them as a function of the simulated hot-versus-mild mixing ratio (see Fig.~\ref{fig:16cygHR}). An observed hardness ratio can then be referenced against this to obtain the hot-versus-mild mixing ratio of the corona of the observed star.

We found that a hard band of 0.6-2.0~keV is suitable for this hardness ratio approach; indeed, the 16 Cyg A and B spectra (Fig.~\ref{fig:16cygspec}) show that the hot 0.5~keV component is mainly responsible for the stellar emission above 0.6~keV for both stars (dotted black lines denoting the spectral contributions of each temperature component). For the soft band, we tried both 0.2-0.6~keV and the broader 0.15-0.6~keV band. We found that both options for the soft band produce very similar estimates for the mild-to-hot mixing ratio, both of which are in line with the ``ground truth'' from the spectral fit. Since the broader 0.15-0.6~keV band yields larger numbers of photons for the faint sample stars and therefore smaller relative uncertainties, we opted to use the 0.15-0.6~keV band as the soft band in our analysis.

\begin{table}
\caption{Results of the spectral analyis of 16 Cyg AC and B on a fixed temperature grid of 0.1 and 0.5~keV, with $1\sigma$ uncertainties on the ``norm'' parameters, which are equivalent to emission measures (see text for details).}
\label{tab:16cyg}
 \begin{small}
 \begin{tabular}{l r r}

 \hline\hline
 Model parameter & 16 Cyg AC & 16 Cyg B \\ \hline
 $kT_1 = 0.086$ keV (1 MK) &  & \\
 norm$_1$ & $(2.48\pm 0.17)\times10^{-5}$  & $(1.17\pm 0.12)\times10^{-5}$ \\
 $kT_2 = 0.34$ keV (4 MK) &  & \\
 norm$_2$ & $(7.07\pm 0.37)\times 10^{-6}$  & $(1.05\pm 0.21)\times 10^{-6}$ \\
 $\chi^2_{red}$ & 1.0  & 0.88 \\[0.2cm]
 \multicolumn{2}{l}{\textit{energy band 0.15-2.0 keV: }} & \\
 $F_X$ [erg/s/cm$^2$]  & $(5.85\pm0.28)\times10^{14}$  & $(2.25\pm0.20)\times10^{14}$ \\
 (from spectral fit) & & \\
 $F_X$ [erg/s/cm$^2$]  & $(5.97\pm0.25)\times10^{14}$  & $(2.02\pm0.19)\times10^{14}$ \\
 (from hardness ratio) & & \\[0.2cm]
 \hline

 \end{tabular}
  \end{small}
\end{table}

To determine X-ray fluxes from the measured count rates for all target stars, we determined the hardness ratio $HR = \frac{H-S}{H+S}$ with $H$ and $S$ being the EPIC count rates in the 0.6-2.0~keV band and the 0.15-0.6~keV band, respectively. We determined those count rates by re-running \verb|edetect_stack| for each target for those specific energy bands. We then used the simulated hardness ratios as explained above to estimate the ratio of mild-to-hot emission measure of each stellar corona and derive the corresponding counts-to-energy conversion factor. The flux was then determined by applying the determined conversion factor to the detected EPIC count rate in the full 0.15-2~keV band. We estimated uncertainties on the flux by performing a Monte Carlo simulation, with samples drawn from Gaussian distributions corresponding to the measured count rates in the bands and their measured uncertainties, and then propagating them through the emission measure estimate to the final fluxes.

\section{Additional figures and tables}

\begin{figure*}
 \includegraphics[width=0.32\textwidth]{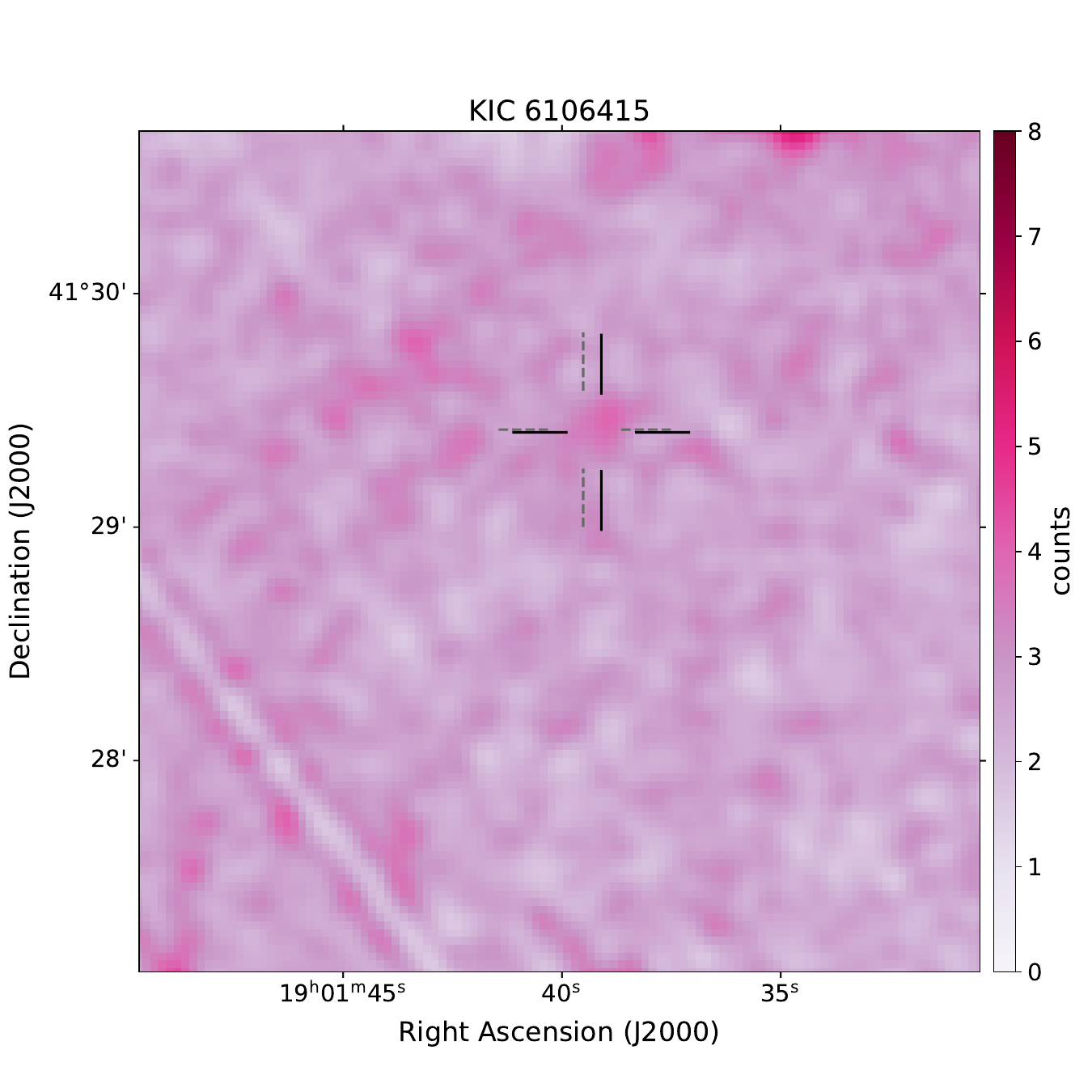}
 \includegraphics[width=0.32\textwidth]{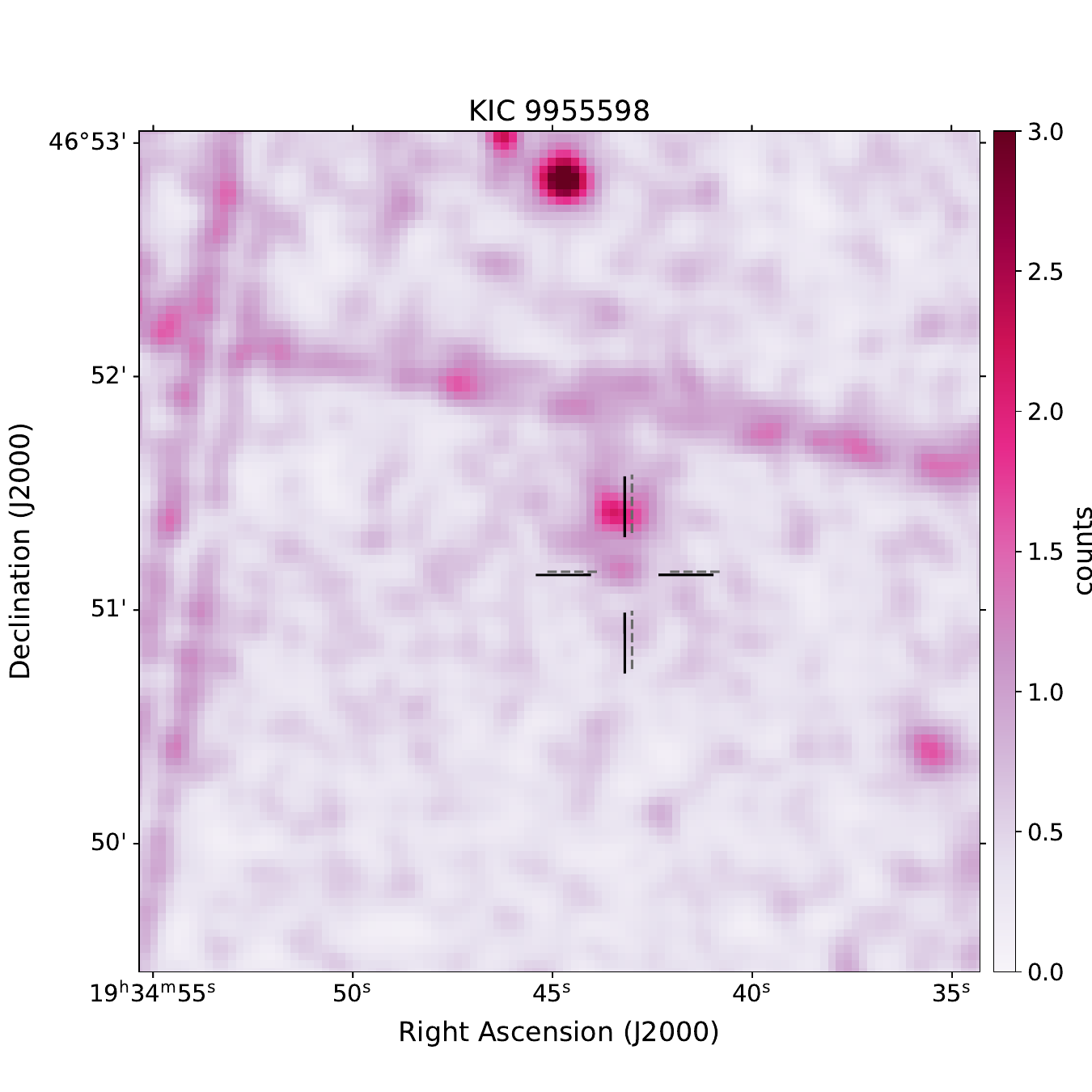}
 \includegraphics[width=0.32\textwidth]{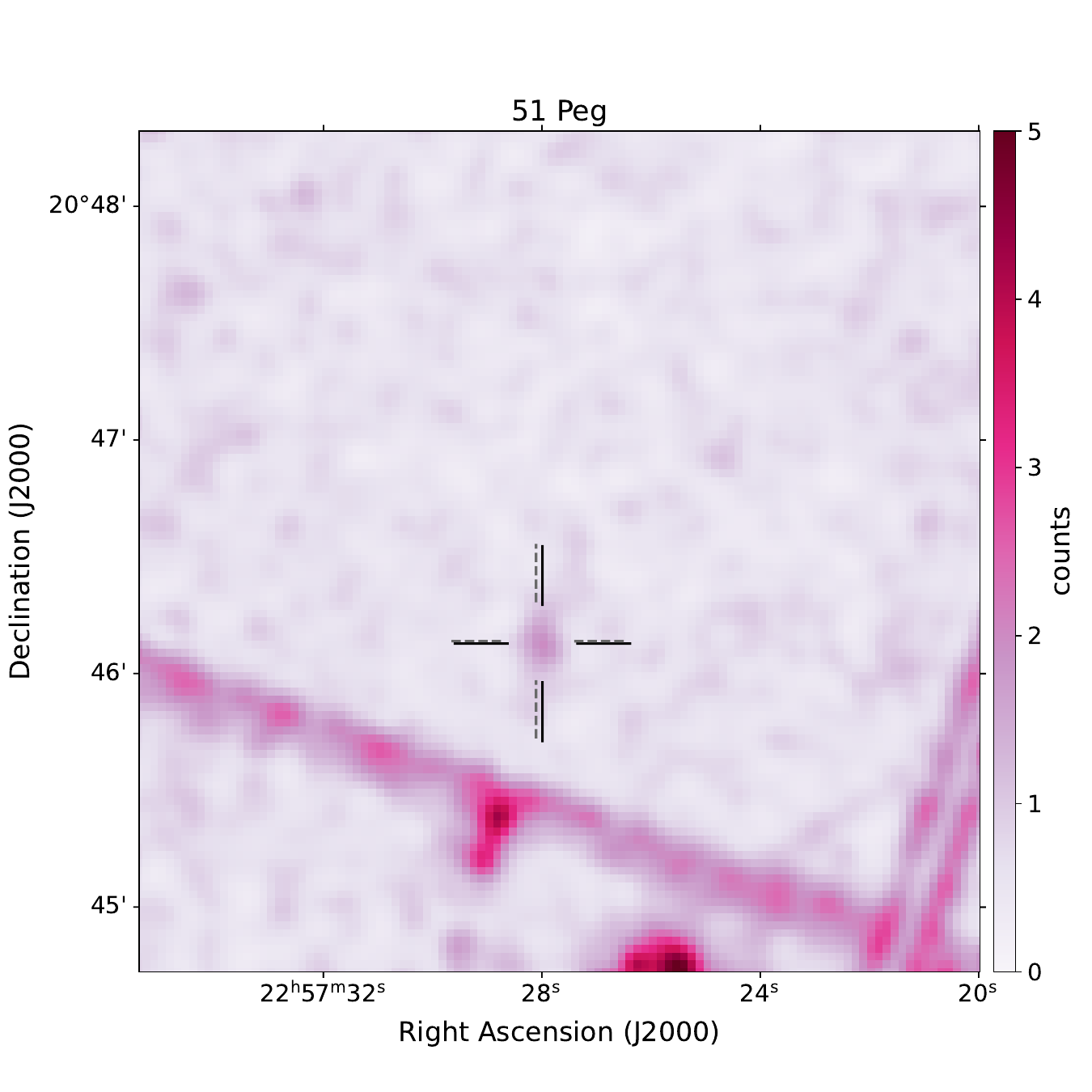}
 \caption{Target stars in addition to 16 Cyg A and B that were successfully detected in X-rays. The reticule with solid lines indicated the detected position of the X-ray source, the dashed one indicates the optical position of the target star from Gaia at the time of the X-ray observation. The shown X-ray data are the combined MOS1, MOS2 and pn photons for the 75-2000~eV band.}
 \label{fig:detections}
\end{figure*}

\begin{figure*}
 \includegraphics[width=0.32\textwidth]{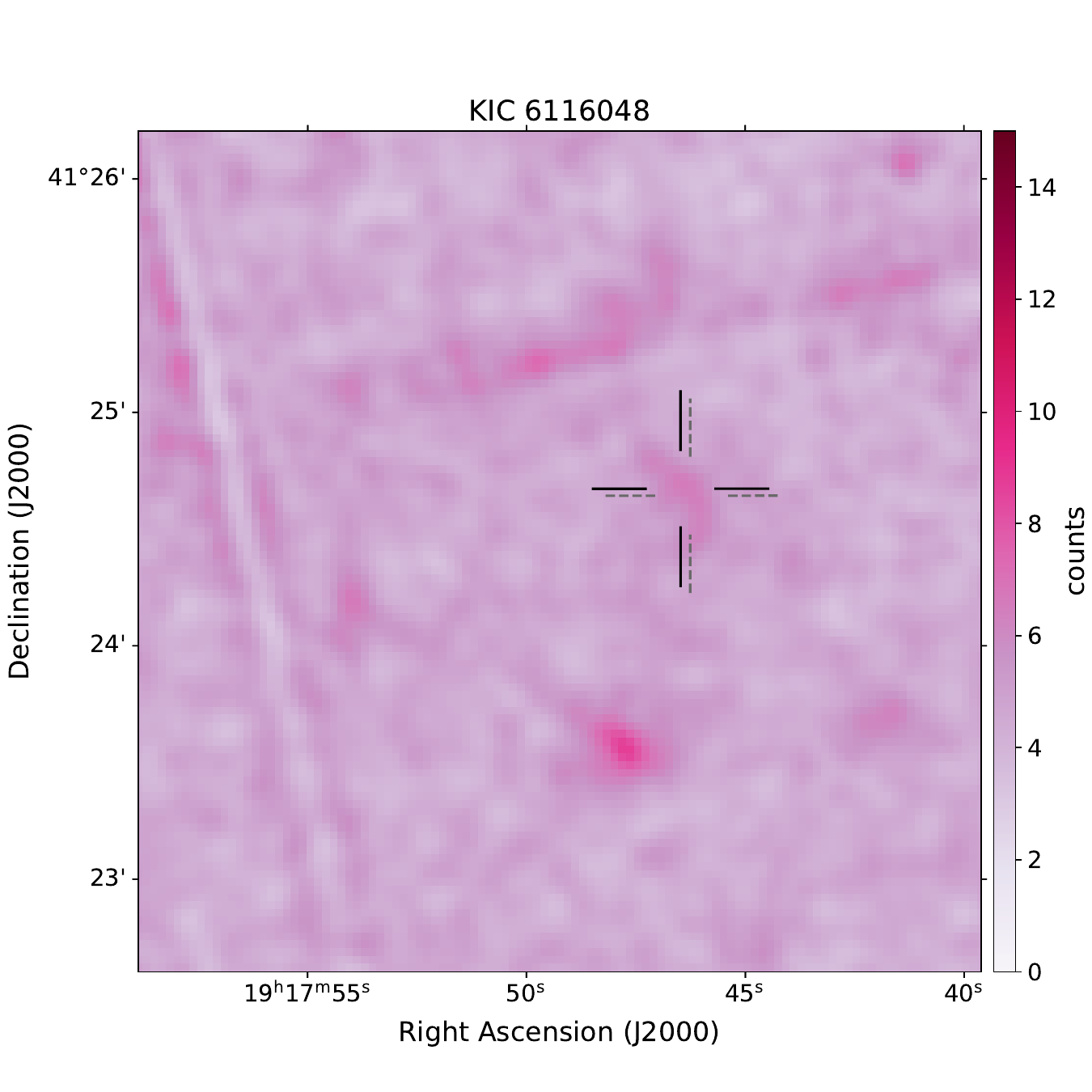}
 \includegraphics[width=0.32\textwidth]{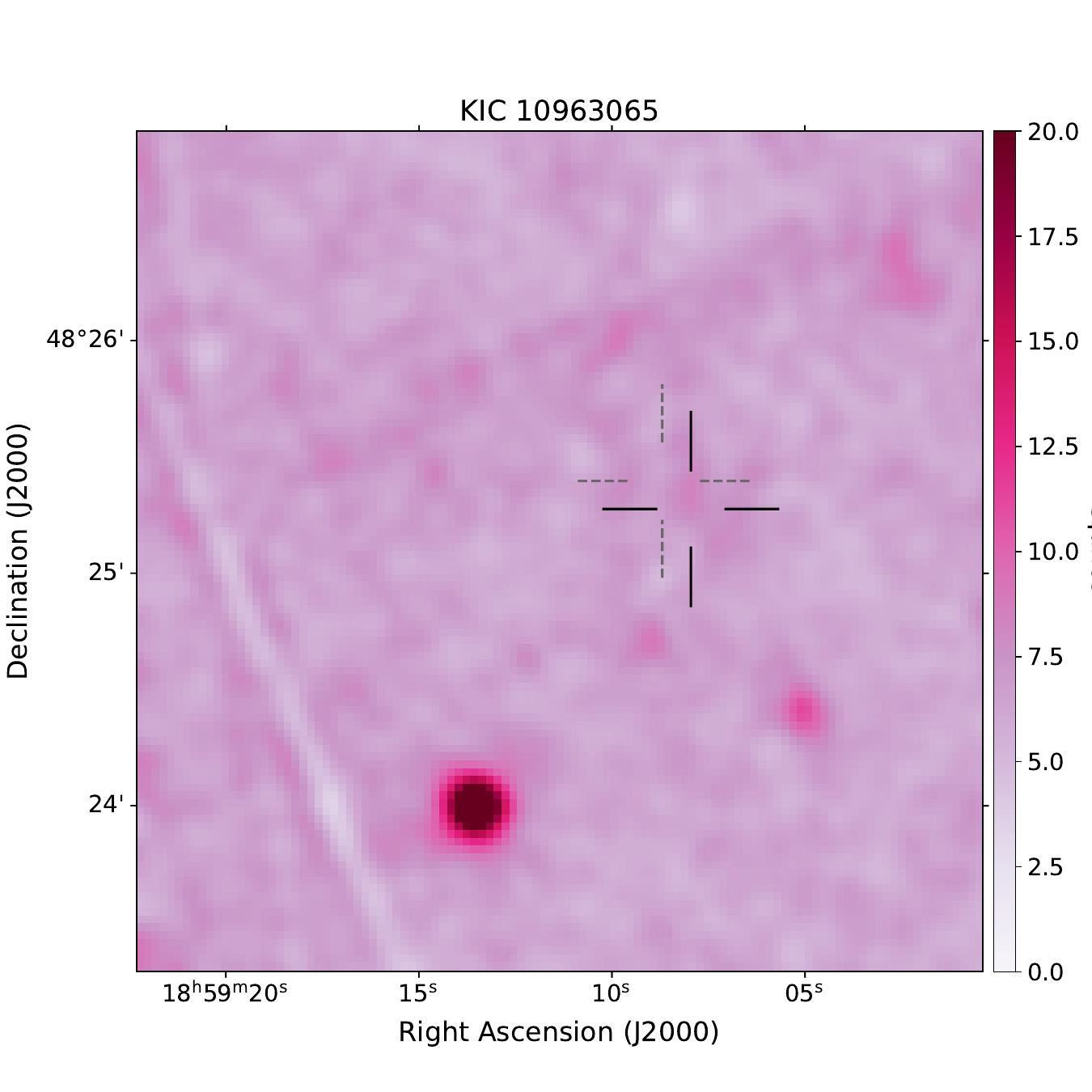}
 \includegraphics[width=0.32\textwidth]{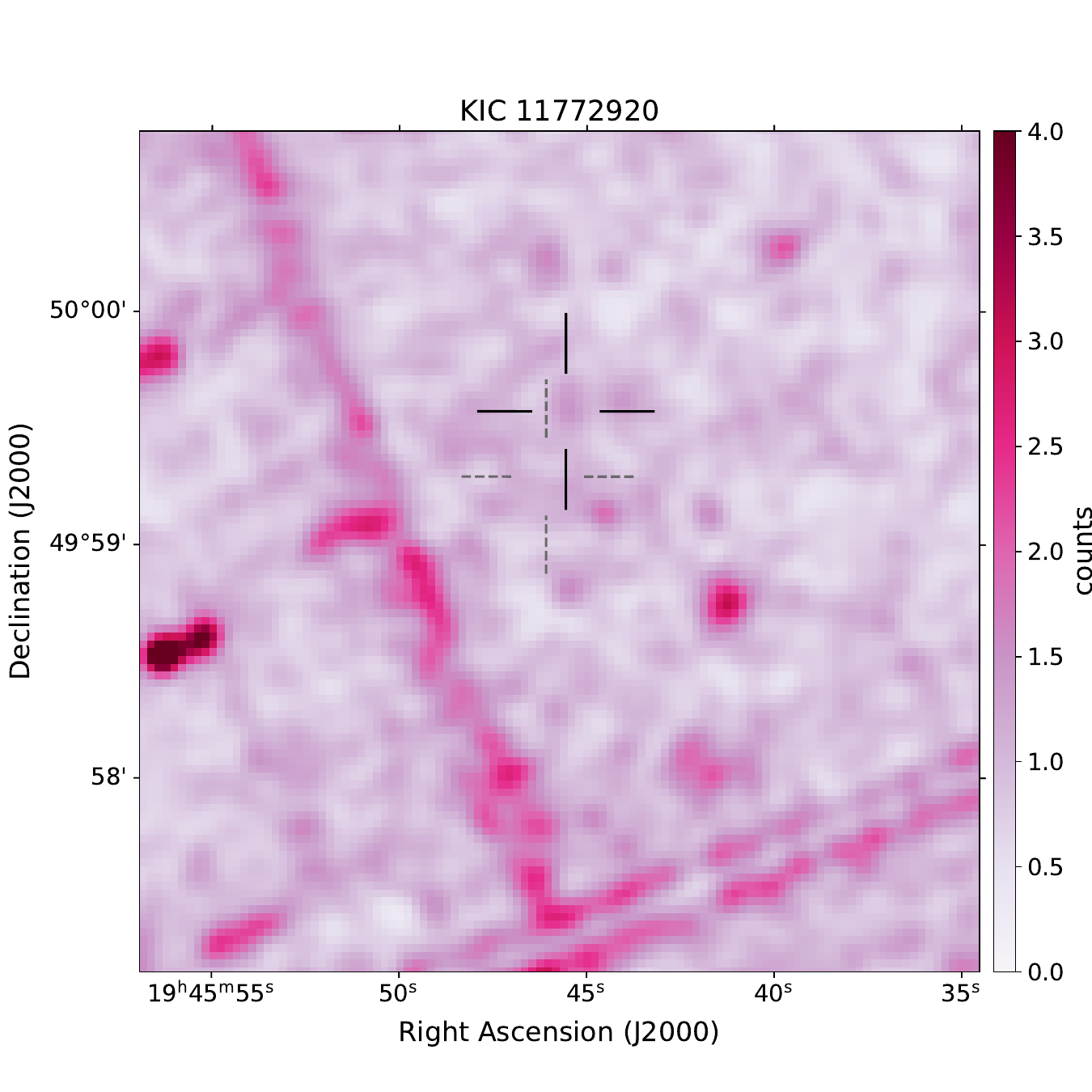}
  \caption{Same as Fig.~\ref{fig:detections}, but for non-detections of target stars. While there is formally a significant photon excess whose centroid coincides with the optical stellar position for KIC~6116048, the extended two-lobe spatial structure is not in agreement with the expected signal for a point source and likely indicates that the excess is unrelated to KIC~6116048. For KIC~10963065 and KIC~11772920, the separation between the detected X-ray excess and the optical stellar position (of ca.\ $9^{\prime\prime}$ and $27^{\prime\prime}$, respectively) is significantly larger than expected for X-ray emission that stems from the same physical object, and is therefore likely unrelated to the target stars.}
 \label{fig:non-detections}
\end{figure*}

\begin{figure*}
 \includegraphics[width=0.32\textwidth]{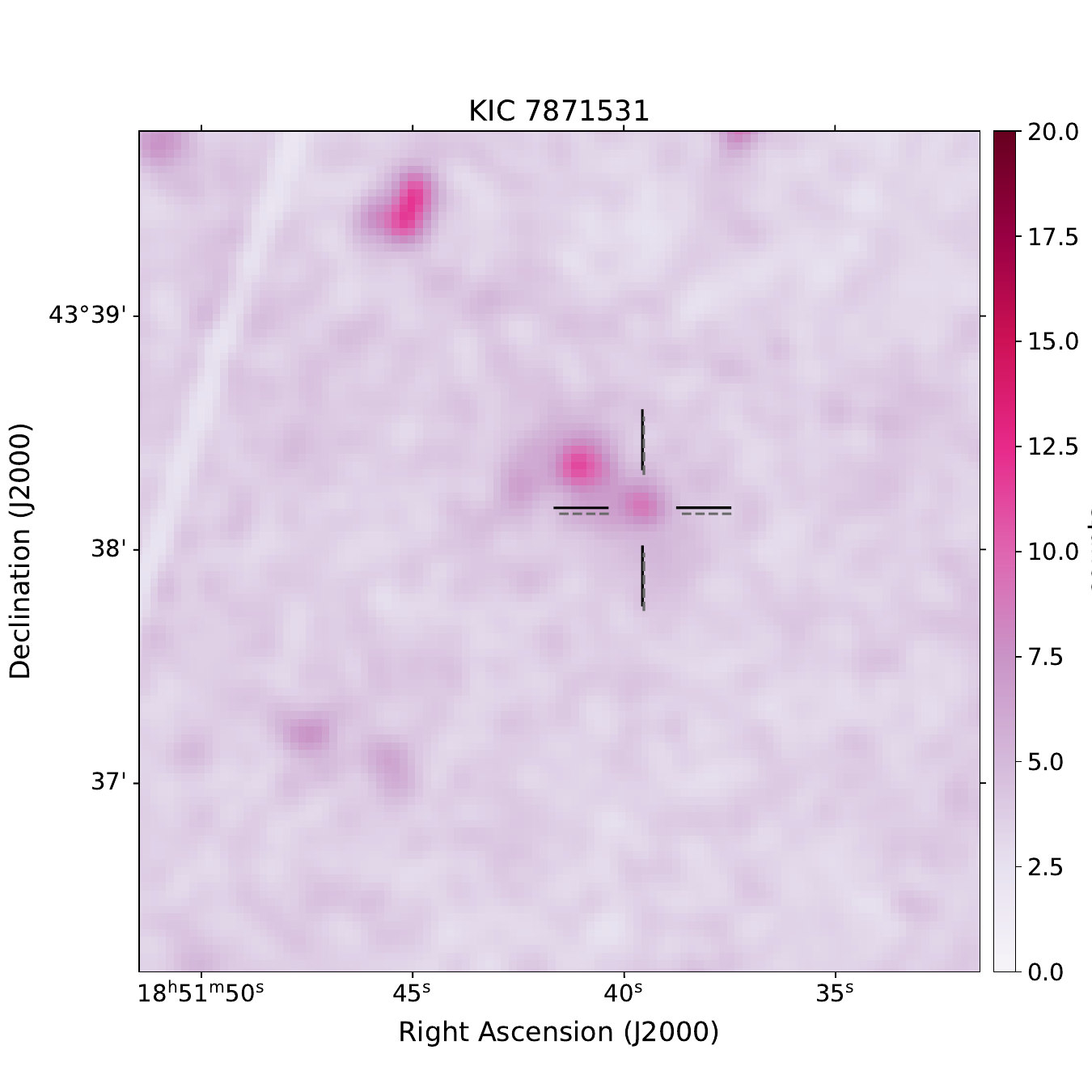}
 \includegraphics[width=0.32\textwidth]{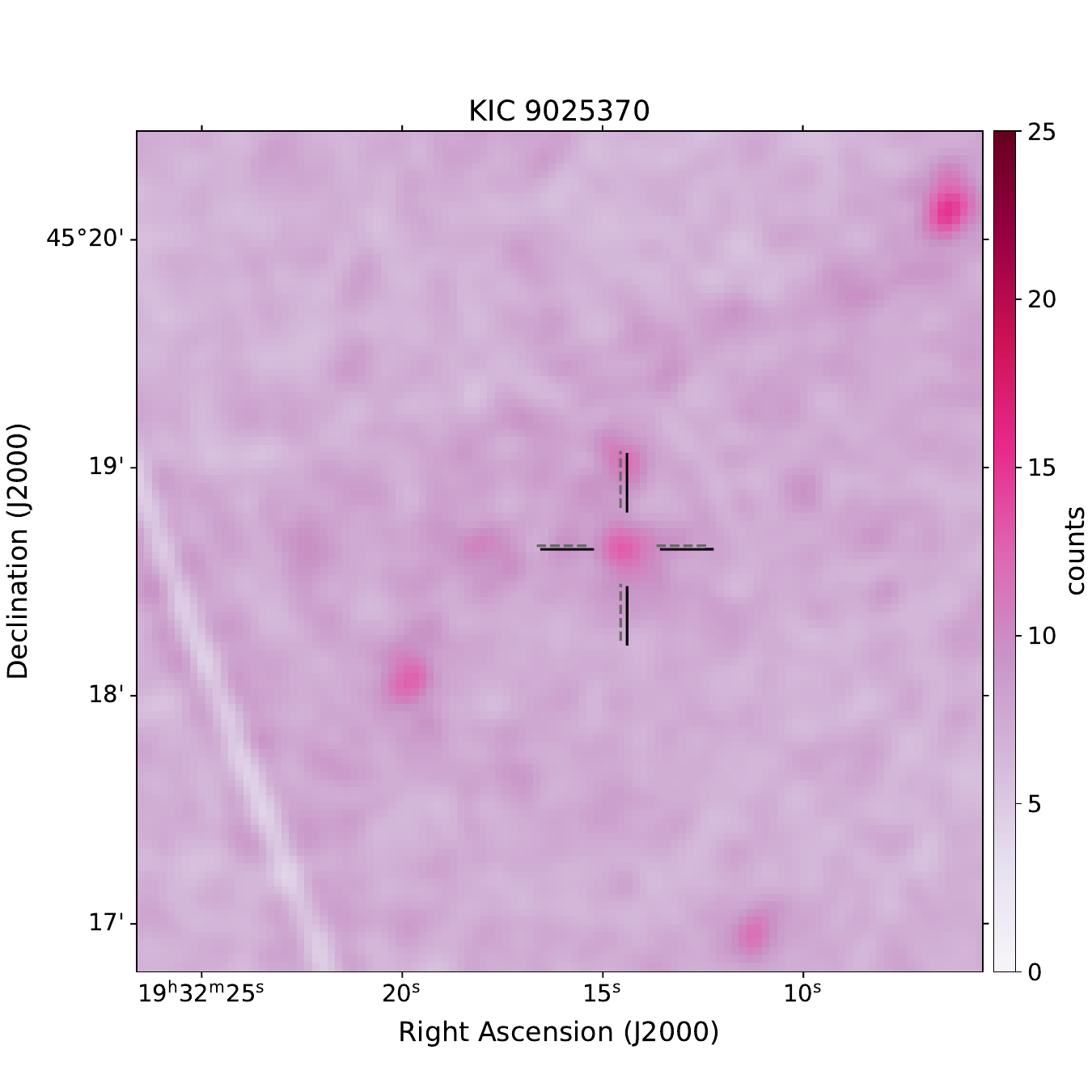}
 \caption{Same as Fig.~\ref{fig:detections}, but for the two stars discarded from the sample.}
 \label{fig:discarded}
\end{figure*}

\begin{table*}
\caption{XMM-Newton observations used in this work.}
\label{tab:obs}
\begin{tabular}{l c r r }
\hline \hline
Target          & XMM Observation ID & Observation start date & Exposure time [s] \\ 
\hline
KIC 10963065    & 0921100101    & 2023-09-10 11:06:00   & 48400  \\
                & 0921100701    & 2023-09-12 13:22:26   & 39600  \\
                & 0921100801    & 2023-09-26 09:29:04   & 50000  \\[0.1cm]
KIC 6106415     & 0921100201    & 2023-09-02 12:31:49   & 45200  \\[0.1cm]
KIC 7871531     & 0921100301    & 2023-10-24 06:30:10   & 53800  \\[0.1cm]
KIC 6116048     & 0921100401    & 2023-09-22 11:05:51   & 45400  \\
                & 0921100901    & 2023-09-24 12:09:42   & 40300  \\[0.1cm]
KIC 11772920    & 0921100501    & 2023-06-24 15:31:39   & 33000  \\
                & 0921101001    & 2023-06-26 19:30:13   & 36900  \\[0.1cm]
KIC 9025370     & 0921100601    & 2023-09-18 10:13:23   & 49550  \\
                & 0921101101    & 2023-09-20 08:11:49   & 56200  \\
                & 0761560501    & 2015-11-17 03:26:50   & 13000  \\[0.1cm]
51 Peg          & 0551020901    & 2008-06-01 11:56:47   & 55321  \\[0.1cm]
16 Cyg AB       & 0823050101    & 2018-11-06 04:51:02   & 73000  \\[0.1cm]
KIC 9955598     & 0761560601    & 2015-10-22 09:45:32   & 27999  \\
\hline

\end{tabular}
\end{table*}

\begin{figure}
\begin{centering}
 \includegraphics[width=0.5\textwidth]{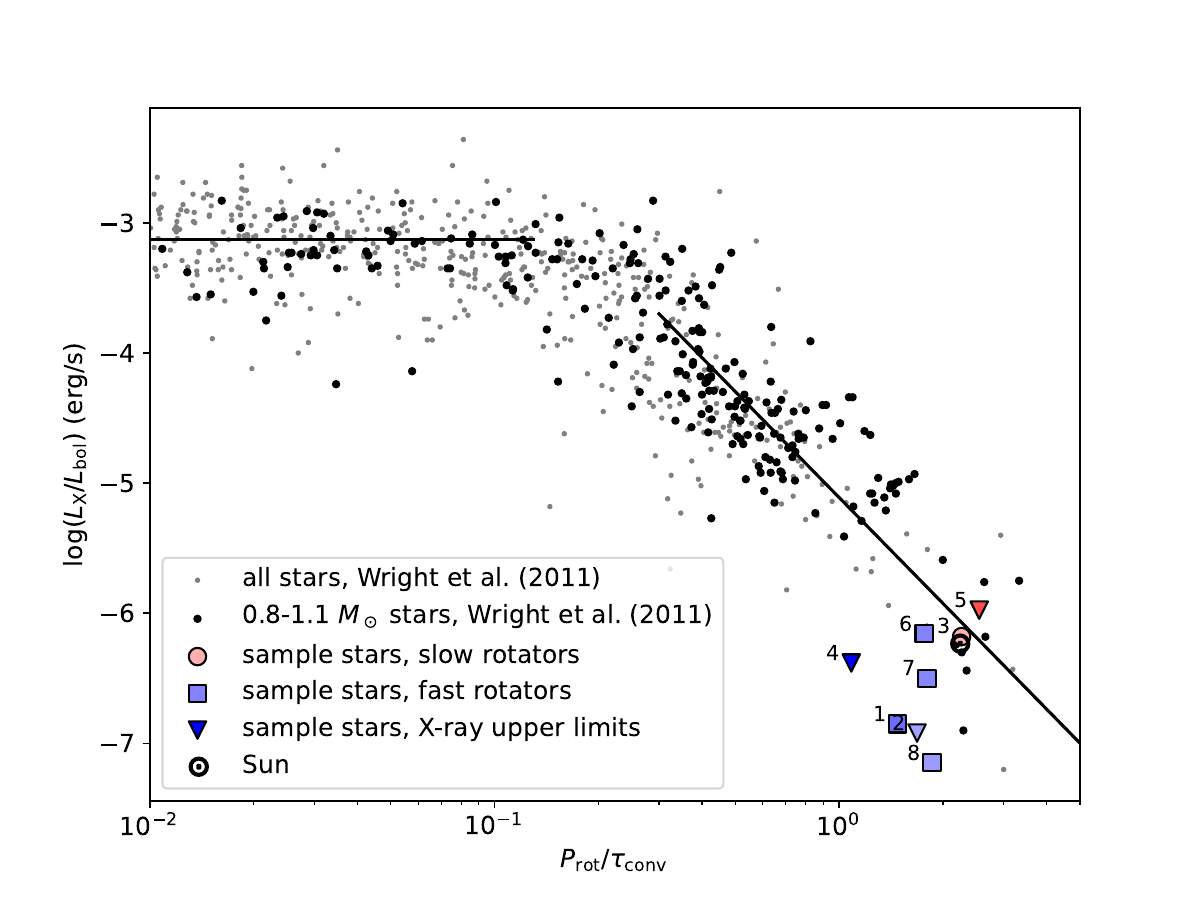}
  \caption{The fractional X-ray luminosity $\log L_X/L_{bol}$ of the target stars in the 0.1-2.4\,keV energy band is shown as a function of the Rossby number, i.e.\ the stellar rotation period divided by the convective turnover time, using the convective turnover estimation method from \cite{Wright2011}. The target stars are shown together with the stellar sample from \cite{Wright2011}. The sample stars cover a wide band of activity levels. Within that band, stars that are rotating faster than their M67 counterparts are shown in successively darker shades of blue; sample stars with stronger over-rotation display lower X-ray activity levels.}
 \label{fig:sample_wright_Rx_Ro}
 \end{centering}
\end{figure}

\begin{figure}
\begin{centering}
 \includegraphics[width=0.5\textwidth]{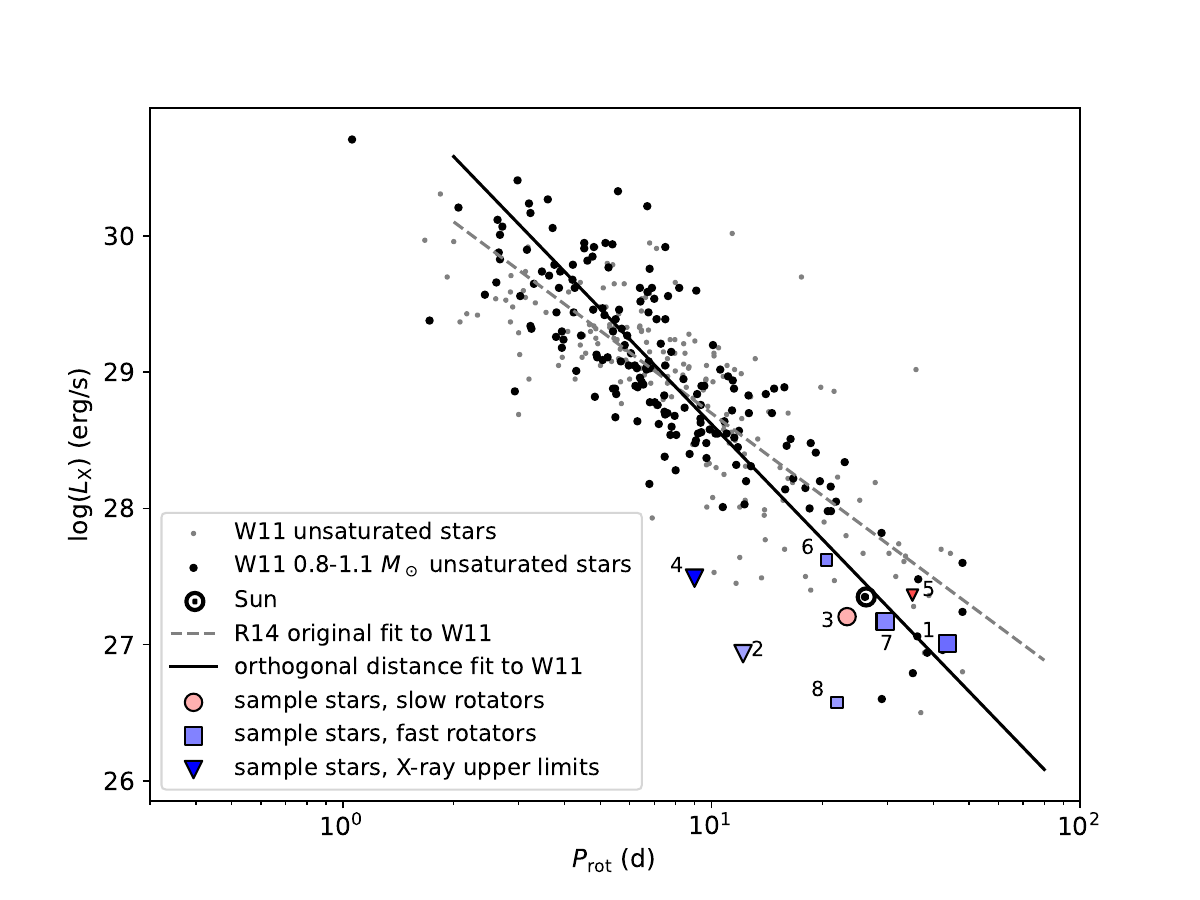}
  \caption{Same as Fig.~\ref{fig:sample_wright}, but with rotation periods from \cite{Reinhold2023} where available for the sample stars shown as large symbols; sample stars without periods from R23 are given with the H21 period as small symbols.}
 \label{fig:rot_act_R23}
 \end{centering}
\end{figure}

\begin{figure}
 \includegraphics[width=0.5\textwidth]{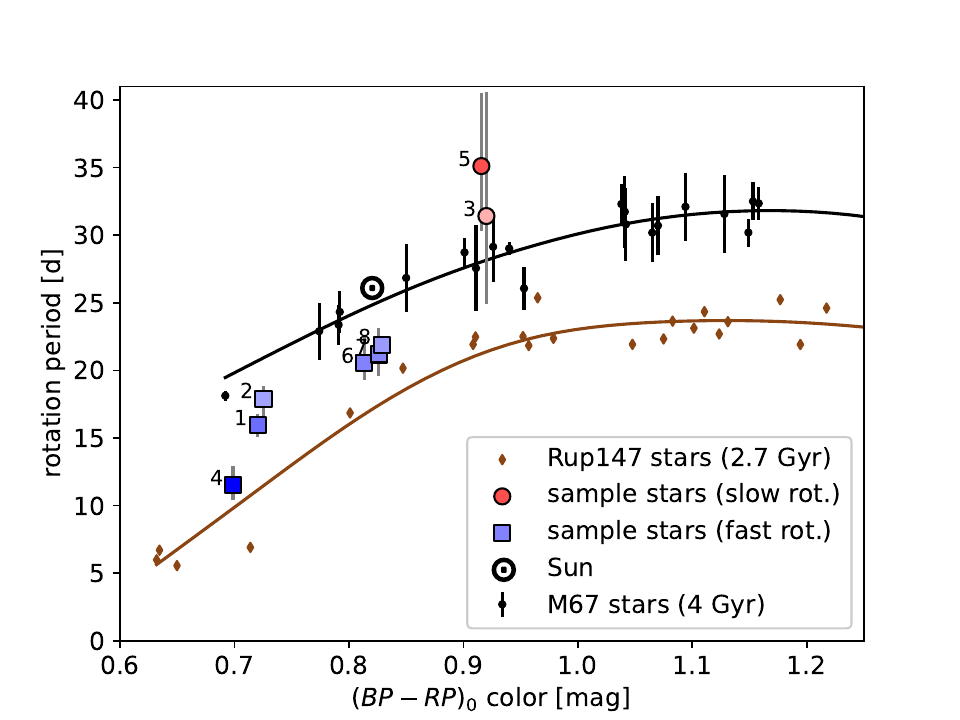}
 \caption{Same as Fig.~\ref{fig:m67rotation}, but with uncertainties on the periods displayed where available (i.e.\ for M67 and our sample stars).}
 \label{fig:periods_zoomed_with_uncertainties}
\end{figure}

\begin{figure}
\begin{centering}
 \includegraphics[width=0.5\textwidth]{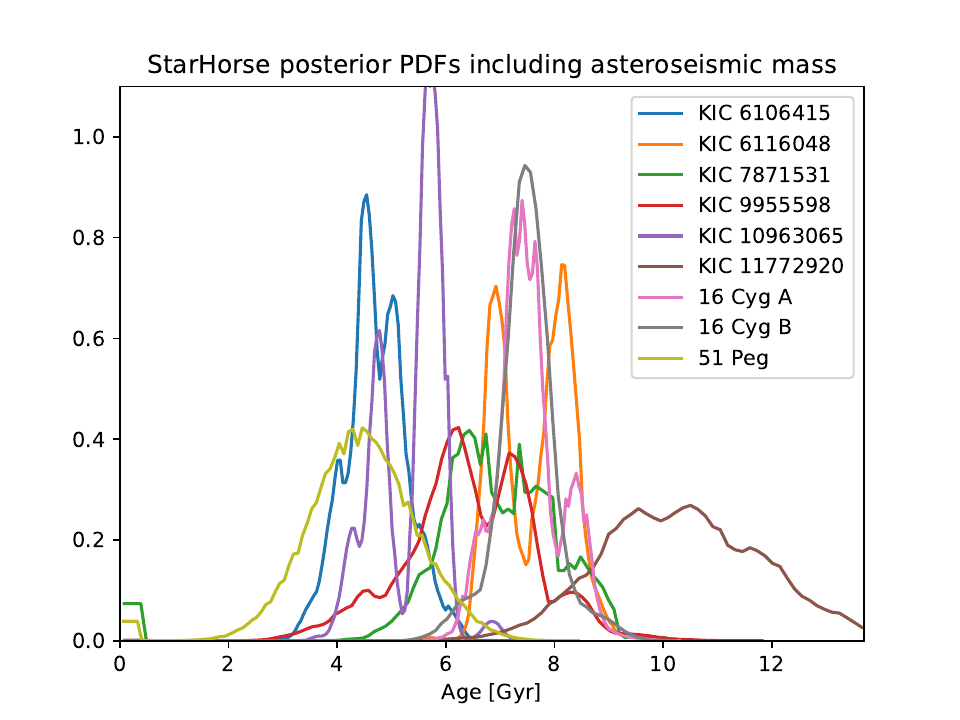}
  \caption{Stellar age posterior probability distributions for our sample stars, derived with {\tt StarHorse}.}
 \label{fig:starhorse_ages}
 \end{centering}
\end{figure}

\clearpage

\onecolumn

\begin{table*}
 \small \caption{Additional properties of the target stars. \label{tab:allresults}}
\begin{tabular}{l r r r r r r r r r r }
\hline\hline
name & index & dist. & radius & mass & age$_{StarHorse}$ & $T_{eff}$ & $\log g$ & $[Fe/H]$ & $P_{rot}$ (R23) & $Ro_{W11}$ \\
 &  &  &  & [M$_\odot$] & [Gyr] & [K] &  &  & [d] &  \\
 & & (Gaia) & (Gaia) & (H21) & (this work) & (H21) & (H21) & (H21) & (R23) & (this work) \\ \hline
KIC 6106415 & 1 & 41.33 & 1.29 & 1.07 & $4.7 \pm0.6$ & $6037 \pm77$ & $4.3\pm0.01$ & $-0.0 \pm0.1$ & $43.8$ & 1.48 \\
KIC 6116048 & 2 & 74.85 & 1.29 & 0.94 & $7.7^{+0.6}_{-0.9}$ & $6033 \pm77$ & $4.25\pm0.01$ & $-0.2 \pm0.1$ & $12.2$ & 1.68 \\
KIC 9955598 & 3 & 68.05 & 0.91 & 0.9 & $6.5 \pm1.1$ & $5457 \pm77$ & $4.5\pm0.01$ & $0.0 \pm0.1$ & $23.3$ & 2.27 \\
KIC 10963065 & 4 & 88.99 & 1.27 & 0.99 & $5.5^{+0.4}_{-0.9}$ & $6140 \pm77$ & $4.28\pm0.01$ & $-0.2 \pm0.1$ & $9.0$ & 1.08 \\
KIC 11772920 & 5 & 70.93 & 0.87 & 0.83 & $10.3^{+1.5}_{-1.4}$ & $5180 \pm180$ & $4.5\pm0.01$ & $-0.1 \pm0.18$ & $--$ & 2.55 \\
16 Cyg A & 6 & 21.12 & 1.3 & 1.05 & $7.5 \pm0.6$ & $5825 \pm50$ & $4.29\pm0.01$ & $0.1 \pm0.03$ & $--$ & 1.77 \\
16 Cyg B & 7 & 21.1 & 1.16 & 0.99 & $7.5 \pm0.5$ & $5750 \pm50$ & $4.35\pm0.01$ & $0.0 \pm0.02$ & $29.6$ & 1.8 \\
51 Peg & 8 & 15.52 & 1.22 & 1.09 & $4.5^{+1.1}_{-1.3}$ & $5758 \pm78$ & $4.32\pm0.08$ & $0.2 \pm0.07$ & $--$ & 1.86 \\
KIC 7871531 & 9 & 67.28 & 1.04 & 0.83 & $4.8 \pm1.3$ & $5501 \pm77$ & $4.48\pm0.01$ & $-0.3 \pm0.1$ & $37.6$ & 2.33 \\
KIC 9025370 & 10 & 83.44 & 1.37 & 0.97 & -- & $5270 \pm180$ & $4.42\pm0.01$ & $-0.1 \pm0.18$ & $19.6$ & 2.07 \\
\hline
\end{tabular}
\end{table*}

\begin{table*}
 \small \caption{X-ray fluxes of the target stars in different energy bands. \label{tab:allresults}}
\begin{tabular}{l r r r r }
\hline\hline
name & index & $F_{X,\,0.1-2.4\,keV}$ & $F_{X,\,0.2-2.0\,keV}$ & $F_{X,\,0.15-2.0\,keV}$ \\
 &  & [erg\,s$^{-1}$\,cm$^{-2}$] & [erg\,s$^{-1}$\,cm$^{-2}$] & [erg\,s$^{-1}$\,cm$^{-2}$] \\
 & & (this work) & (this work) & (this work) \\ \hline
KIC 6106415 & 1 & $(4.99 \pm 2.16) \times 10^{-15}$ & $(3.28 \pm 0.93) \times 10^{-15}$ & $(4.16 \pm 1.53) \times 10^{-15}$ \\
KIC 6116048 & 2 & $<1.28\times 10^{-15}$ & $<5.12\times 10^{-16}$ & $<9.0\times 10^{-16}$ \\
KIC 9955598 & 3 & $(2.89 \pm 1.49) \times 10^{-15}$ & $(1.81 \pm 0.65) \times 10^{-15}$ & $(2.39 \pm 1.07) \times 10^{-15}$ \\
KIC 10963065 & 4 & $<3.24\times 10^{-15}$ & $<1.3\times 10^{-15}$ & $<2.28\times 10^{-15}$ \\
KIC 11772920 & 5 & $<3.83\times 10^{-15}$ & $<1.54\times 10^{-15}$ & $<2.7\times 10^{-15}$ \\
16 Cyg A & 6 & $(7.81 \pm 0.36) \times 10^{-14}$ & $(4.07 \pm 0.14) \times 10^{-14}$ & $(5.97 \pm 0.25) \times 10^{-14}$ \\
16 Cyg B & 7 & $(2.77 \pm 0.27) \times 10^{-14}$ & $(1.25 \pm 0.1) \times 10^{-14}$ & $(2.02 \pm 0.19) \times 10^{-14}$ \\
51 Peg & 8 & $(1.3 \pm 0.32) \times 10^{-14}$ & $(5.25 \pm 1.23) \times 10^{-15}$ & $(9.19 \pm 2.19) \times 10^{-15}$ \\
KIC 7871531 & 9 & $(1.7 \pm 0.3) \times 10^{-14}$ & $(9.1 \pm 1.13) \times 10^{-15}$ & $(1.31 \pm 0.21) \times 10^{-14}$ \\
KIC 9025370 & 10 & $(1.15 \pm 0.2) \times 10^{-14}$ & $(5.24 \pm 0.75) \times 10^{-15}$ & $(8.42 \pm 1.37) \times 10^{-15}$ \\
\hline
\end{tabular}
\end{table*}

\clearpage
\twocolumn

\end{appendix}

\end{document}